\documentstyle[referee,psfig]{aa}
\input epsf         
\begin{document}

\title{X-ray He-like ions diagnostics : New Computations for 
Photoionized 
Plasmas: I. preliminary considerations}

\author{S\'everine Coup\'e\inst{1}, 
Olivier Godet\inst{2}, Anne-Marie Dumont\inst{1}, 
Suzy Collin\inst{1}}

\offprints{Olivier Godet (Olivier.Godet@cesr.fr)}

\institute{$^1$LUTH, Observatoire de Paris, Section de
Meudon, Place Jules Janssen, F-92195 Meudon Cedex\\
$^2$CESR,  9 av. du Colonnel Roche, 31028 Toulouse
Cedex 4, France}

\date{Received : / Accepted : }

\titlerunning{X-ray He-like ions diagnostics : New Computations
for Photoionized Plasmas}
\authorrunning{S. Coup\'e et al.}

\abstract{Using the new version of the photoionization code Titan designed 
for plane-parallel photoionized thick hot media, which is 
unprecedented from the point of view of line transfer,
 we have undertaken a systematic 
study of the influence of different parameters on the He-like and 
H-like emission of a medium photoionized by an X-ray source. We 
explain why in modelling the emitting medium it is important to 
solve in a self-consistent way the thermal and ionization equilibria 
and to take into account the interconnection between the different 
ions. 
 We insist on the influence of the column density on the 
He-like ion emission, via stratification of ion species, temperature 
gradient, resonance trapping and continuum absorption, and we show 
that misleading conclusions can be deduced if it is neglected. In 
particular a given column density of an He-like ion can lead
 to 
a large range of total column densities and ionization parameters. We 
show also that there is a non-model-dependent relation between an 
ion column density and its corresponding temperature, and 
that the ion column density  cannot exceed a maximum value for a 
given ionization parameter. We give the equivalent widths of the sum 
of the He-like triplets and the triplet intensity ratios $G$ and $R$, 
for the most important He-like ions, for a range of density, column 
density, and ionization parameter, in the case of constant 
density media.  We show in particular that the line intensities 
from a given ion 
can be accounted for, either by small values of both the column density 
and of the ionization parameter, or by large values of both quantities, 
and it is necessary to take into account several ions to disentangle 
these possibilities.  We show also that a ``pure recombination spectrum" almost never 
exists in a photoionized medium: either it is thin, and resonance lines are 
formed by radiative excitation, or it is thick, and free-bound 
absorption destroys the resonance photons as they undergo resonant 
diffusion. Consequently, the $G$ ratio is much smaller than the pure 
recombination ratio for a small value of the total column 
density, and it exceeds
the recombination ratio for large values of the total column density 
and of the ionization parameter.

}

\maketitle

\section{Introduction}

With the great sensitivity and resolution of the new generation of 
X-ray missions XMM-Newton and Chandra, detailed spectra of several 
types of objects have been obtained in the soft X-ray range, showing 
tens of emission lines which can be used as diagnostics of the 
physical state of the emitting regions. For instance it is now 
possible to separate the He-like ions lines of the $n=2$ complex in 
many objects. These lines are used to determine the electronic 
temperature and the density of the observed region. 

Gabriel \& Jordan (1969, 1972, 1973) showed that, in collisional plasmas, 
the resonance, intercombination and forbidden lines ratios of He-like 
ions have very interesting properties. The ratio, called $R$, of the 
forbidden line $z$ ($1s^2\ ^1S\ - 2s\ ^3S$) and the intercombination 
lines $y$ and $x$ ($1s^2\ ^1S\ - 2p\ ^3P^o_{1,2}$) :
\begin{equation}
R = \frac{z}{x+y}
\end{equation} 
 is sensitive to the 
electronic density $n_e$. 
The $G$ ratio of the forbidden plus 
intercombination lines over 
the resonant line $w$ ($1s^2\ ^1S\ - 2p\ ^1P^o$)~:
\begin{equation}
G = \frac{z+x+y}{w}
\end{equation} 
is sensitive to the electronic temperature.

The $R$ and $G$ ratios have therefore been extensively used to 
determine electron
densities and temperatures of hot collisional plasmas.
Recently, theoretical calculations have extended these ratios to 
photoionized and hybrid plasmas for extragalactic objects like the 
warm absorbers which are supposed to give rise to the O VII and O 
VIII edges in Active Galactic Nuclei (AGNs) (Porquet \& Dubau 2000 
for hybrid and photoionized plasmas, and Bautista \& Kallman 2001 for 
collisional and photoionized plasmas).  In these papers however, 
``photoionized conditions"  mean only that the spectrum is due to 
recombination, radiative cascades and collisional excitation, but 
photoionization and photoexcitation are not taken 
into account. 
Porquet et al. (2001) have introduced a term of photoexcitation in 
collisional plasmas, for the study of the spectra of late-type 
dominated coronae and O-B stars. The radiation field was a diluted 
photospheric black body of a few 10000K, so it had an influence only 
on the visible and UV transitions, but not on the resonance line 
excitation. 

If the X-ray emitting plasma is photoionized, the irradiating 
continuum {\it necessarily excite the resonance and the 
subordinate permitted lines}. The importance of photoexcitation 
on the population levels and on the line ratios has 
also started to be taken into account.
This was stressed for the first time by Sako et al. (2000) in their 
analysis of the spectrum of the Seyfert 2 galaxy Mrk 3. Kinkhabwala 
et al. (2002) showed that it can account for the observed ratios of 
NVI and OVII lines in the spectrum of the Seyfert 2 galaxy NGC 1068, 
where the observed $R$ ratio is larger and the $G$ ratio is smaller 
than in a pure recombination case.

However, even these last ``photoionized" computations  are not 
completely self-consistent, as the temperature is a free parameter 
and the thermal equilibrium is not solved.
Kinkhabwala et al. (2002) assumed a multi-zone model (with a geometry 
adapted to the case of Seyfert 2 galaxies) 
 where each ion is present in a different layer, and where the 
temperature is set a priori, as determined by the observed width of 
the corresponding radiative recombination continuum (RRC). 
The ion column densities are determined by fitting the X-ray 
spectrum (Ogle et al. 2003 use the same method to fit the  X-ray 
observations of NGC 1068). The code PHOTO of Kinkhabwala, which is now 
implemented in the XSPEC package, takes well into account the geometry of 
the emitting medium, but use different approximations which are not
valid for a moderately thick medium,  mainly: photoelectric absorption of only 
the ion under study is taken into account, the medium is optically 
thin to the emitted photons, the thermal equilibrium is not 
consistently computed with the ionization equilibrium, and a very approximate 
line escape probability is used.

Finally, in all computations made with photoionized codes, even the 
most sophisticated ones like Cloudy (Ferland et al. 1998) or XSTAR
 (Kallman \& Krolik 1995, Kallman \& Bautista 2001), the transfer of the lines is not performed, or it 
is performed with an ``escape probability" approximation. 
The transfer of the continuum also is generally done through an 
``outward only" approximation. These approximations are not valid for 
a thick medium, as shown by comparing the results of these 
approximations with full line transfer performed with our code Titan 
(see Dumont et al. 2000, Dumont et al. 2003). This precludes 
the use of these codes for column densities larger than  10$^{23}$ 
cm$^{-2}$. We will see that neglecting the possibility of large 
column densities can lead to misleading interpretations of the 
observations.

\bigskip

The medium observed in emission or in absorption in the X-ray range 
span a large range of physical conditions:

\begin{itemize} 

\item In Seyfert 1, a few lines and edges broadened by large Doppler 
motions are observed in emission over an intense continuum: resonance 
lines of OVII (the different components of the triplet cannot be 
resolved) and OVIII, the fluorescent FeK line, and  photoionization 
edges of the same ions. 
 All these features are due to ``reflection" in the atmosphere of the 
accretion disc irradiated by the X-ray continuum (see for 
instance Nayakshin et al. 2000, Ballantyne et al. 2001, R{\' o}{\. 
z}a{\' n}ska et al. 2002). This reflected spectrum is formed in a few 
Thomson thicknesses below the surface, i.e. up to a column 
density of about 10$^{25}$ cm$^{-2}$.
Much narrower spectral features appear in absorption against 
the underlying X-ray continuum. 
They are due to a medium photoionized by the central X-ray continuum 
and located on its line of sight, called the Warm Absorber (WA).
 The WA also emits a few lines and continua from H-like and He-like 
ions, principally OVII and OVIII, which are difficult to detect. They 
can be observed as P Cygni profiles when the S/N ratio is high, like 
in NGC 3783 (Kaspi et al. 
2002). Though the location and the density of the WA is 
controversial, it is clearly a relatively dense medium, with a 
density $n_H$ larger than 10$^{7}$ cm$^{-3}$, located 
close to the black hole, at a distance $R$ of the order of that of 
the Broad Line Region, and having a column density $CD$ of the order 
of 10$^{21-23}$ cm$^{-2}$.
There are even suggestions that the warm absorber could be located closer 
to the black hole, with a larger column density (10$^{24}$ cm$^{-2}$, 
see Murray \& Chiang 1995, Green \& Mathur 1996), and is 
denser, since one should preserve the value of the ionization 
parameter $\propto n^{-1}R^{-2}$ which determines the ionization 
state. 
Porquet et al. (2000) argue also that the density should be larger 
than 10$^{9}$ cm$^{-3}$ to avoid the emission of too strong forbidden 
optical coronal lines of FeX and FeXIV which are not observed.

\item Splendid X-ray spectra of a few Seyfert 2 galaxies have been 
obtained recently with Chandra and XMM-Newton (Ogle et al. 2003, Sako 
et al. 2000,  Kinkhabwala et al. 2002). They are most probably 
produced by the reflecting medium invoked in the Unified Scheme of 
Seyfert 1/Seyfert 2 (Antonucci \& Miller 1985). This medium is 
identified with the external part of the WA, but it is seen in 
emission, as the intense central continuum in Seyfert 2 is hidden to 
our view by an obscuring torus, according to the Unified Scheme. This 
medium is also photoionized by the central continuum, but it is 
located further away than the WA, as the lines are relatively narrow 
(between the broad and the narrow line widths). Consequently the 
density should be smaller, probably not larger than 10$^{7}$ cm$^{-3}$. 

\item There are other types of ``photoionized" objects where high 
resolution X-ray spectra showing the He-like and H-like lines have 
been observed with XMM-Newton and Chandra, allowing a study of the 
physical conditions in the emission region: X-ray binaries (Kallman 
et al. 2003, Jimenez-Garate et al. 2002, Schulz \& Brandt, 2002) and 
cataclysmic variables (Mukai et al. 2003). The emissive medium is 
either an accretion column or simply the accretion disc atmosphere. 
The density is large, of the order of 10$^{11}$ cm$^{-3}$, and the 
column density can reach 10$^{23}$ cm$^{-2}$, though the geometry is 
clearly quite complex.

\end{itemize}

We have therefore decided to build a grid of models encompassing all 
those photoionized cases where X-ray lines are observed. We give here 
some preliminary results. Our model atom is still not very elaborate 
(see below), so the  precision in the line fluxes and line 
ratios is only of the order of 50$\%$, but they can be used  at least 
to get a hint of the physical parameters corresponding to the 
emission regions, and of the trends of the spectrum when these 
parameters vary.
We insist, in particular, on the influence of the column 
density on the He-like ion emission, as our calculations are reliable 
using our line transfer method for large column densities, which were  
not properly taken into account previously, when they were 
greater than 10$^{22}$ cm$^{-2}$.

\section{The computational method}

\subsection{The photoionization-transfer code}

In this work we use our code Titan specially designed for an optically 
thick warm or hot medium. Titan is a photoionization stationary code 
first
described in Dumont et al. (2000). The atomic data have recently been 
improved (Coup\'e 2002). It was used by R{\' o}{\. z}a{\' n}ska et 
al. (2003) to fit the spectrum of the Warm Absorber of Ton S180 with 
a slab of constant pressure. In the present paper, we use another 
version of Titan implemented very recently (Dumont et al. 2003) where 
the line and the continuum transfer is solved with the  Accelerated 
Lambda Iteration (ALI) method (see for instance the review of Hubeny 
2001). This powerful method ensures that even the most optically 
thick lines are accurately computed.   
  This is a big advantage with respect to the other photoionization 
codes, where the transfer of the lines (and in some cases also the 
transfer of the continuum) is replaced by an integral method called 
the ``escape probability approximation".  
Dumont et al. (2003) have indeed shown that this approximation leads 
to inaccuracies of the computed line fluxes which can reach {\it  one 
order of magnitude}, in the case of an optically thick medium, in 
particular for intense lines. In our computation, all line fluxes are 
computed with an accuracy better than $1\%$.
 
A plane parallel slab of gas is irradiated by an incident spectrum, 
whose frequency-integrated flux is equal to $F$. We call ``ionization 
parameter" $\xi = 4\pi\ F/n_H$ erg cm s$^{-1}$, where $n_H$ is the 
hydrogen number density. Some authors prefer to use an ionization 
parameter
integrated in the X-ray range or from 1 Rydberg to infinity. 

 The gas composition includes the 10 most abundant elements (H, He, 
C, N, O, Ne, Mg, Si, S, Fe) and all their ion species, i.e. 102 
ions. H-like, He-like, Li-like, O~IV and O~V include a multi-level 
description up to $n=5$ for H-like and Li-like ions, $n=4$ for 
He-like ions and $n=3$ for O IV and O V, where $n$ is the principal 
quantum number, which ensures a correct computation of
line losses in the zones where He-like ions are present.
 Interlocking between excited levels is included. 
Other ions are treated more roughly.

Titan includes all relevant physical processes from each level: 
photoionization and radiative and dielectronic recombination, 
ionization by high energy photons, fluorescence and Auger processes, 
charge exchanges, collisional ionization and 3-body 
recombination, radiative and collisional excitation/deexcitation. All 
induced processes are also included. The populations of each 
level are computed solving the set of ionization equations coupled 
with the set of statistical equations describing the excitation 
equilibrium, taking into account radiative and collisional 
ionizations from all levels and recombinations onto all levels, 
radiative and collisional excitations and deexcitations for all 
transitions.

Titan solves the ionization equilibrium of all the ion species of 
each element, the thermal equilibrium, the statistical equilibrium of 
all the levels of each ion, and the transfer of the lines and of the 
continuum. It gives as output the ionization and temperature 
structures, and the reflected and outward spectra.  In particular the 
energy balance is achieved both {\it locally} (which leads to the 
value of the equilibrium 
temperature as a function of the position) with a precision better 
than 0.01$\%$, and {\it globally} (Titan 
checks that the integrated flux entering the slab is equal to that 
leaving the slab from both sides with a precision better than $1\%$). 
It was shown in Dumont et al. (2003) that such a global equilibrium 
cannot be achieved with an escape probability approximation with a 
precision better than about $10\%$ if the Thomson thickness of the 
slab is of the order of unity, and this has a strong impact on the 
line spectrum.

\subsection{The He-like atom model}

The He-like atom has been extensively described in the literature 
(Gabriel \& Jordan 1969; Blumenthal et al. 1972, Mewe \& Schrijver 
1978, Pradhan et al. 1981, Pradhan \& Shull 1981; Pradhan 1982, 1985, 
Porquet \& Dubau 2000, Bautista \& Kallman 2001).
Our present model for He-like ions is made up of all terms for $n=2$, 
and 4 super-levels gathering the singlet and triplet levels for n=3 
and 4. Recombinations on the upper levels $n>4$ are taken into 
account. Satellite lines are not included.
 We are aware that these 11 levels plus a continuum are not 
sufficient to get a very accurate description of the atom. 
However we have checked that in the case of a completely thin medium 
we get the correct ratios $G$ and $R$ within an accuracy of $50\%$, for a 
pure recombination spectrum as well as for pure collisional cases.  
In the future we intend to implement a larger number of levels and to 
add doubly excited states.

 The wavelengths of the lines are taken from the compilation of the 
National Institute of Standards and Technology. We use the transition 
probabilities $A$ computed or collected by Porquet \& Dubau 
(2000). The TOPbase data provides the photoionization cross sections 
(Cunto et al. 1993). The radiative recombination rate coefficients 
are obtained by means of the Milne relation, removing the resonances 
by interpolation between points on opposite sides of the resonances, 
while total radiative recombination rate coefficients are from Verner 
\& Ferland (1996). The dielectronic rate coefficients are from 
Pequignot et al., (private communication 1986). We adopt the 
collisional data of Zhang \& Sampson (1987), with the data for $T=0$ 
from Porquet (private communication).  We use electron impact ionizations rates from Arnaud \& Raymond
 (1992) and private communication from Arnaud.
Three-body recombination rates are obtained by means of detailed 
balance.

\section{Results: the thermal and ionization structure}

We have run a series of models with a density varying from 10$^{7}$ 
cm$^{-3}$ to 10$^{12}$ cm$^{-3}$, a range of column densities from 
10$^{18}$ cm$^{-2}$ to 3 10$^{25}$ cm$^{-2}$, and a range of 
ionization parameters from 10 to 3000.  We assume cosmic abundances 
with respect to hydrogen (Allen 1973): He: 0.085; C: 3.3 10$^{-4}$; 
N: 9.1 10$^{-5}$; O: 6.6 10$^{-4}$; Ne: 8.3 10$^{-5}$; Mg: 2.5 
10$^{-5}$; Si: 3.3 10$^{-5}$; S: 1.6 10$^{-4}$; Fe: 3.2 10$^{-5}$.

 The incident spectrum is a power law, $F_\nu \propto \nu^{-1}$, 
extending from 0.1 eV to 100 keV (we will call it the ``standard 
continuum"). 
It is not a very good representation of the continuum in AGN, so we 
have also made a few runs with a more realistic continuum: it is a 
power law $F_\nu \propto \nu^{-0.7}$ extending from 13.6 eV 
to 100 keV  (we call it the ``AGN continuum").  

Actually these models do not take into account all possible 
situations. 

\begin{itemize}

\item The 
element abundances have an obvious influence on the emitted 
spectrum, since the intensities of optically thin lines is 
proportional to the abundance and those of optically thick lines 
($\tau_{0}\ge 1/a$, where $\tau_{0}$ is the optical thickness at the 
line center, and $a$ is the damping factor) are proportional to the 
square root of the abundance. 

\item The spectral distribution of the incident continuum also has an 
influence on the temperature and on the ionization state. For 
instance, at large densities ($\ge 10^{13}$ cm$^{-3}$), free-free 
heating becomes important, owing to the
extension of our continuum in the infrared range. More important, a 
steeper or flatter continuum in the soft X-ray range would induce a 
different stratification in ionization. It is interesting thus to 
rely on the X-ray ionization parameter, because the ionization 
potentials of the species of interest for us are in the range 400 eV 
to 3 keV. If we define $\xi_{x}$ as given by the integration of the 
flux above 1 keV, one finds that $\xi_{x}= 0.33\xi$ with our standard 
continuum. For the AGN continuum, one gets $\xi_{x}({\rm 
AGN})=2.4\xi_{x}({\rm standard})$ for the same value of the flux at 1 
keV. Since in the following we want to compare the results obtained 
with the standard continuum to those obtained with the AGN continuum, 
we will call this ionization parameter ``equivalent $\xi$", 
such as  $\xi_{\rm eq}=2\xi$, for the AGN continuum.

\item We have chosen to deal with a constant density inside the slab, 
while a medium with a constant pressure would lead to a different 
ionization and thermal structure (in particular smaller column 
densities of highly ionized species, see Dumont et al. 2002).  

\item Finally we do not take into account the possible existence of 
microturbulence, considering only thermal broadening.
 If the emitting medium is dilute and continuous, one should indeed consider 
the possibility of a large velocity gradient, like in a wind. The 
intrinsic width of the lines would then be larger, and their optical 
thicknesses would be smaller. It would have an influence on 
relatively thin models, where radiative excitation of the lines is 
still important (cf. below), because there is more available 
continuum radiation. It would also have an influence on the escape of 
resonance lines in the case of thick models. Of course, if the 
medium is clumpy and made of clouds with large velocities with 
respect to each other, or with a velocity gradient, this 
``macroturbulence" has to be taken into account in the spectrum, but 
does not intervene in the line transfer (like in the Broad Line 
Region of AGN). 
\end{itemize}

The influence of all these parameters will be studied in a subsequent 
paper.

To understand the results on the intensity ratios, it is first 
necessary to study the ionization and thermal state of the emitting 
medium.

One should realize that a photoionized medium is not uniform when it 
is optically thick for the absorption of the soft X-ray continuum. 
The temperature decreases and the medium becomes less ionized towards 
the non-illuminated side, as the incident continuum is more absorbed. 
A thick slab is thus divided into a ``hot skin" below the surface, 
where the temperature is almost constant and the elements are highly 
ionized, an ``intermediate" zone where the temperature decreases by 
about one order of magnitude and the elements are less ionized, and a 
``cold'' zone, where the elements are in low ionization states. If 
the column density of the slab is not large enough, it will be made 
only of the hot skin, or possibly of the hot skin and a fraction of 
the intermediate zone. The thickness of the hot skin and the 
temperature at the surface increase with the value of the ionization 
parameter. The decrease of the temperature through the slab is large 
for large values of the ionization parameter. It is therefore {\it 
not possible to define a unique temperature, even for a given ion 
species}. Moreover, if lines are formed by recombination, the 
abundance of the upper ion giving rise to the lines can vary strongly 
in the region where the ion corresponding to the transition is 
present.
  
\begin{figure}
\begin{center}
\psfig{figure=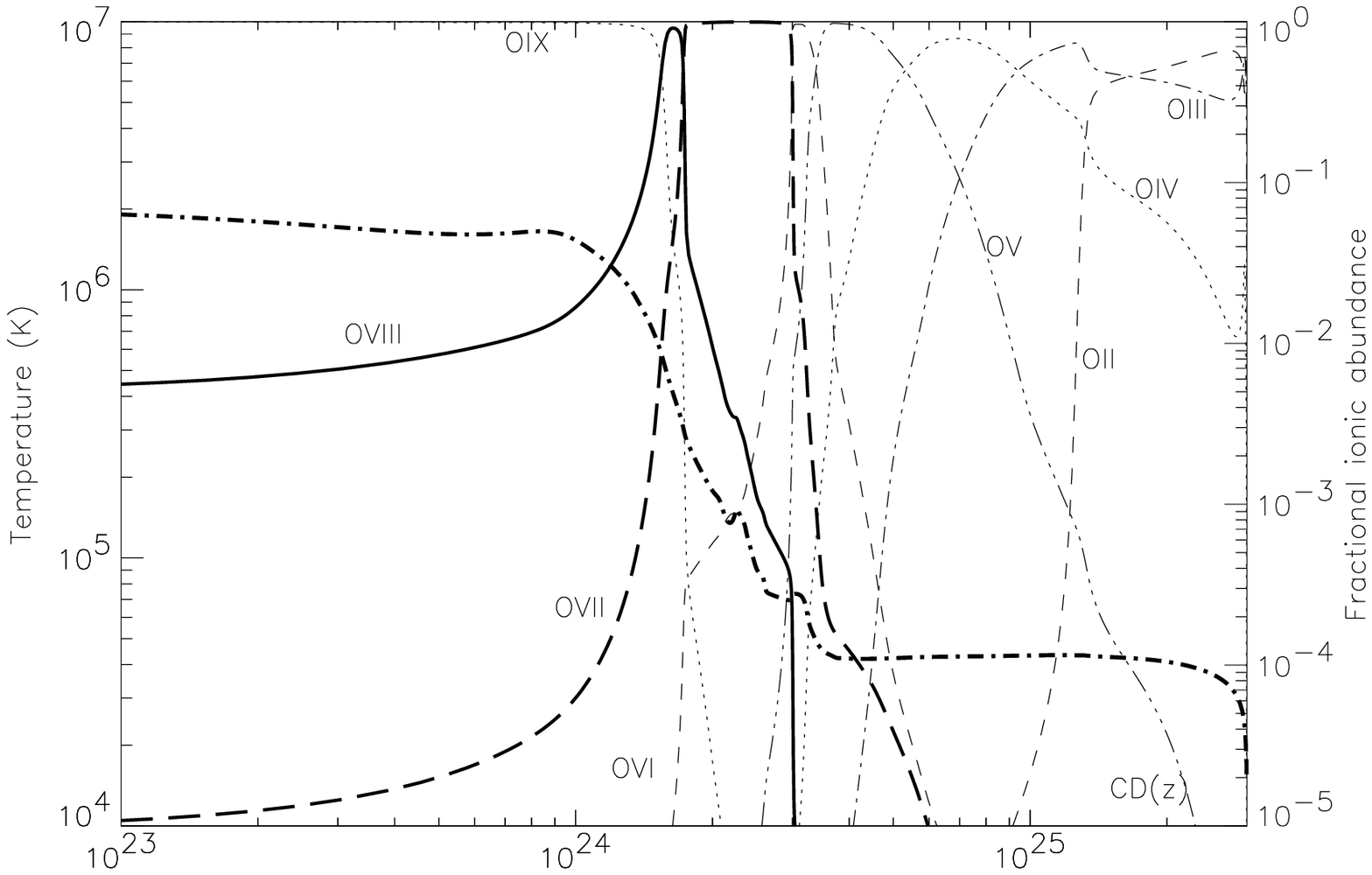,width=7.5cm,height=7.cm}
\vspace{-0.5cm}
 \psfig{figure=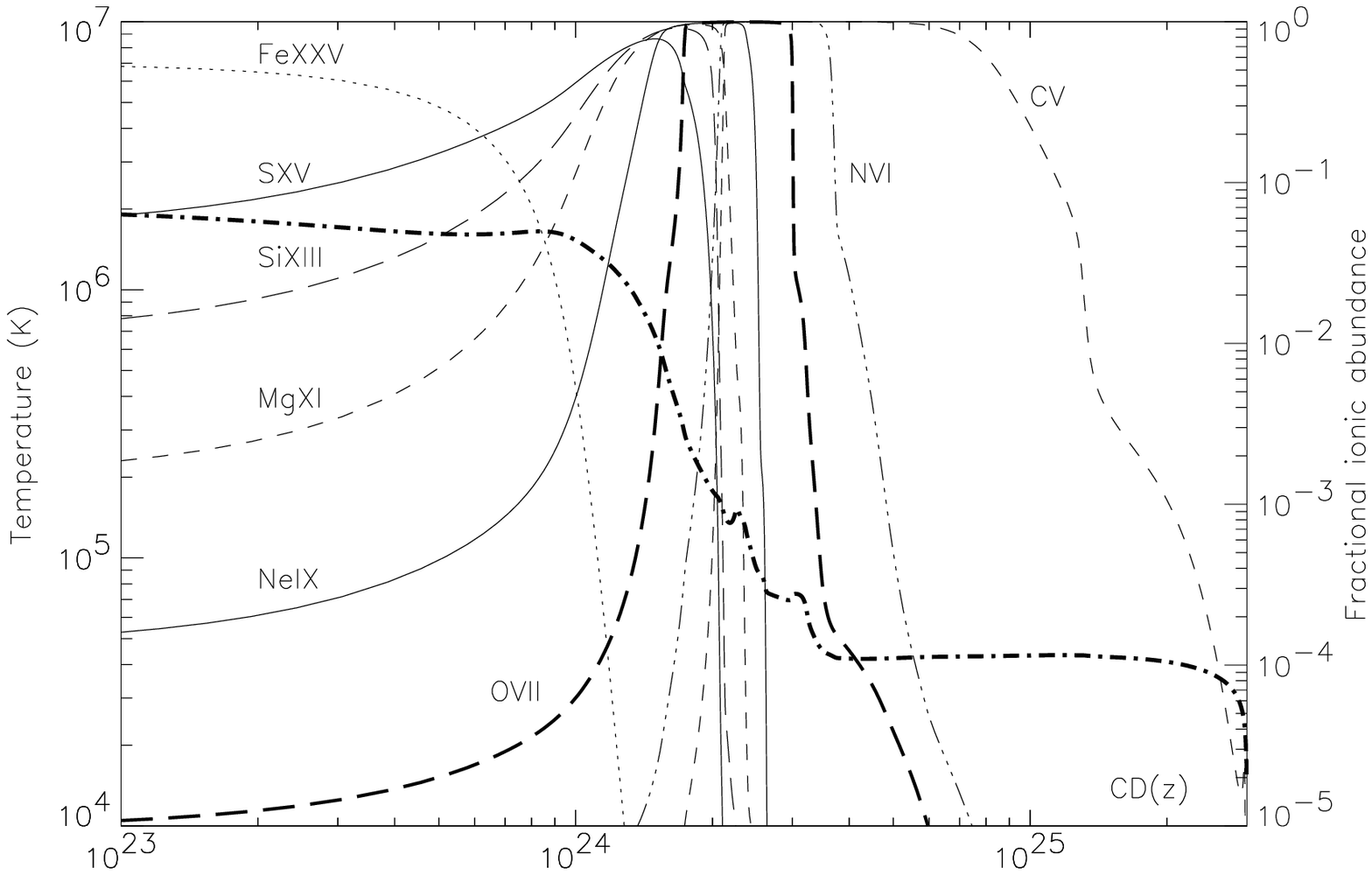,width=7.5cm,height=7.cm} 
\vspace{-0.5cm}
\psfig{figure=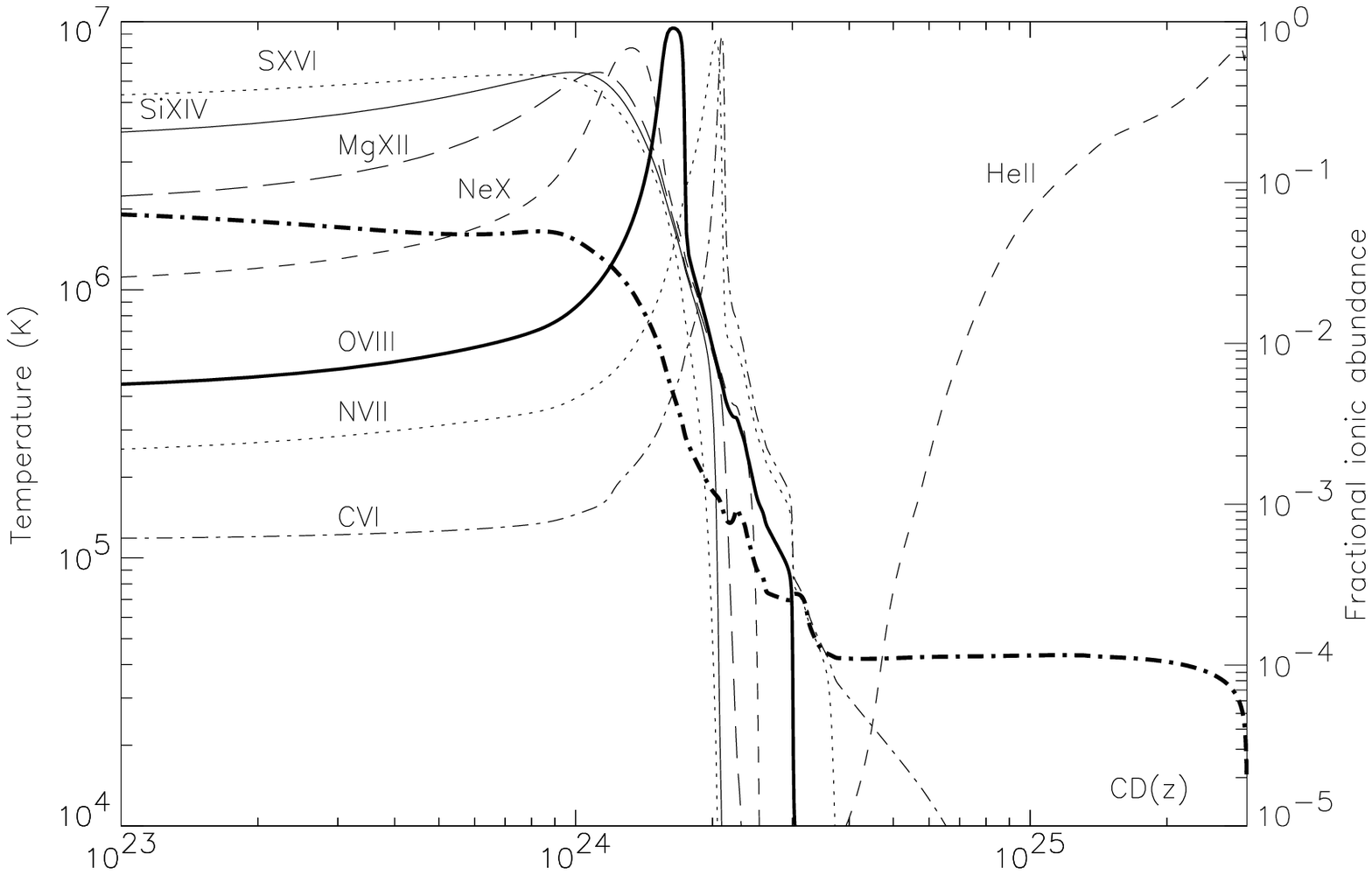,width=7.5cm,height=7.cm}
\caption{Temperature and fractional ion abundances versus $CD(z)$
 for a model with 
$CD= 3\times 10^{25}$ cm$^{-2}$, $n_{H}=10^{12}$ 
cm$^{-3}$, 
  and an ionization parameter of $\xi=1000$, photoionized by the 
standard 
  continuum (Model 1). The thick dot-dashed line corresponds 
to the temperature. The top figure shows the fractional 
abundances of
  the oxygen ions, the middle one the fractional abundances of the 
  He-like ions, and the bottom one, the fractional abundances of 
  the H-like ions.}
  \label{fig1}
\end{center}
\end{figure}

\begin{figure}
\begin{center}
\psfig{figure=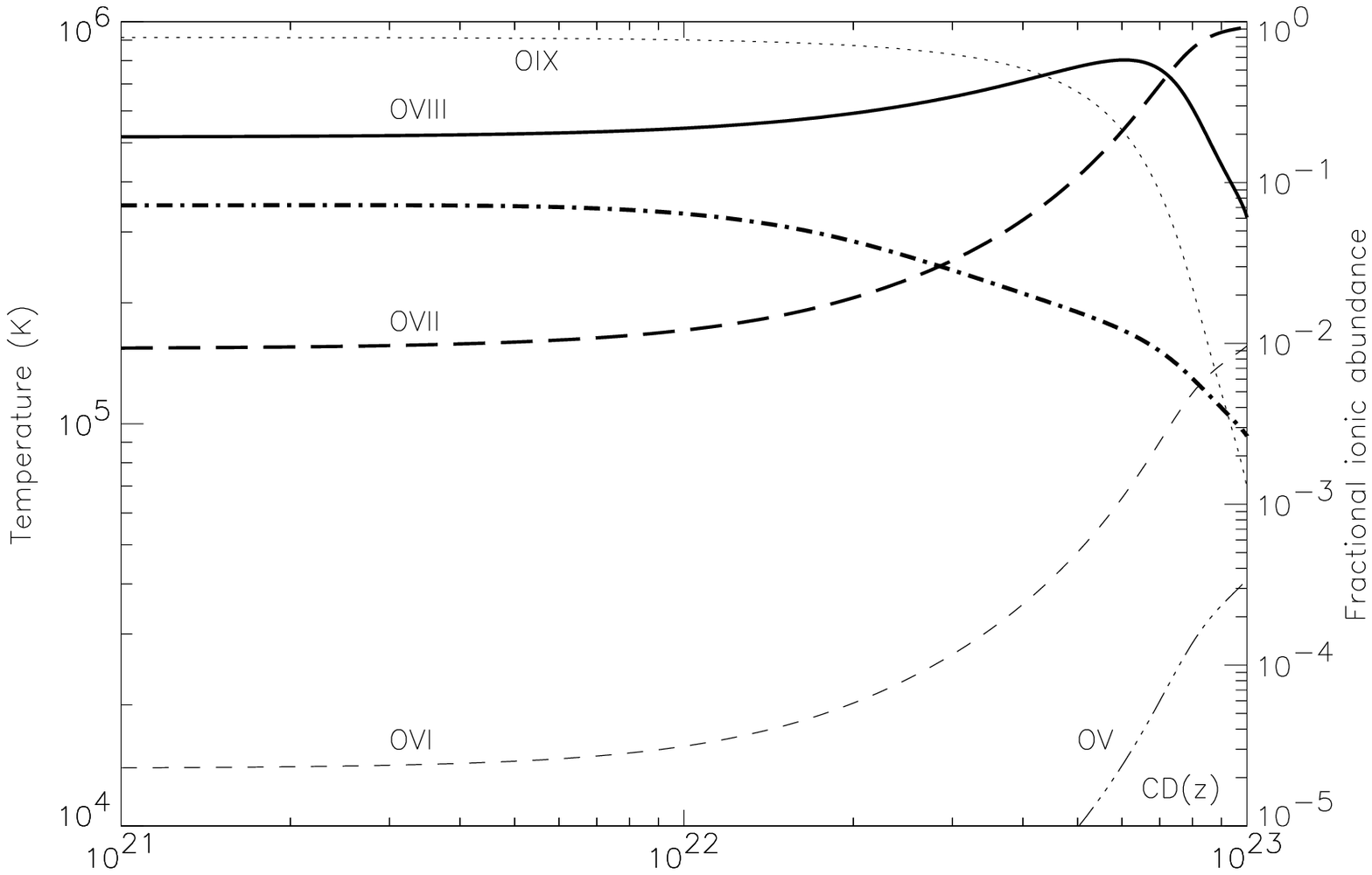,width=7.5cm,height=7.cm}
\vspace{-0.5cm}
\psfig{figure=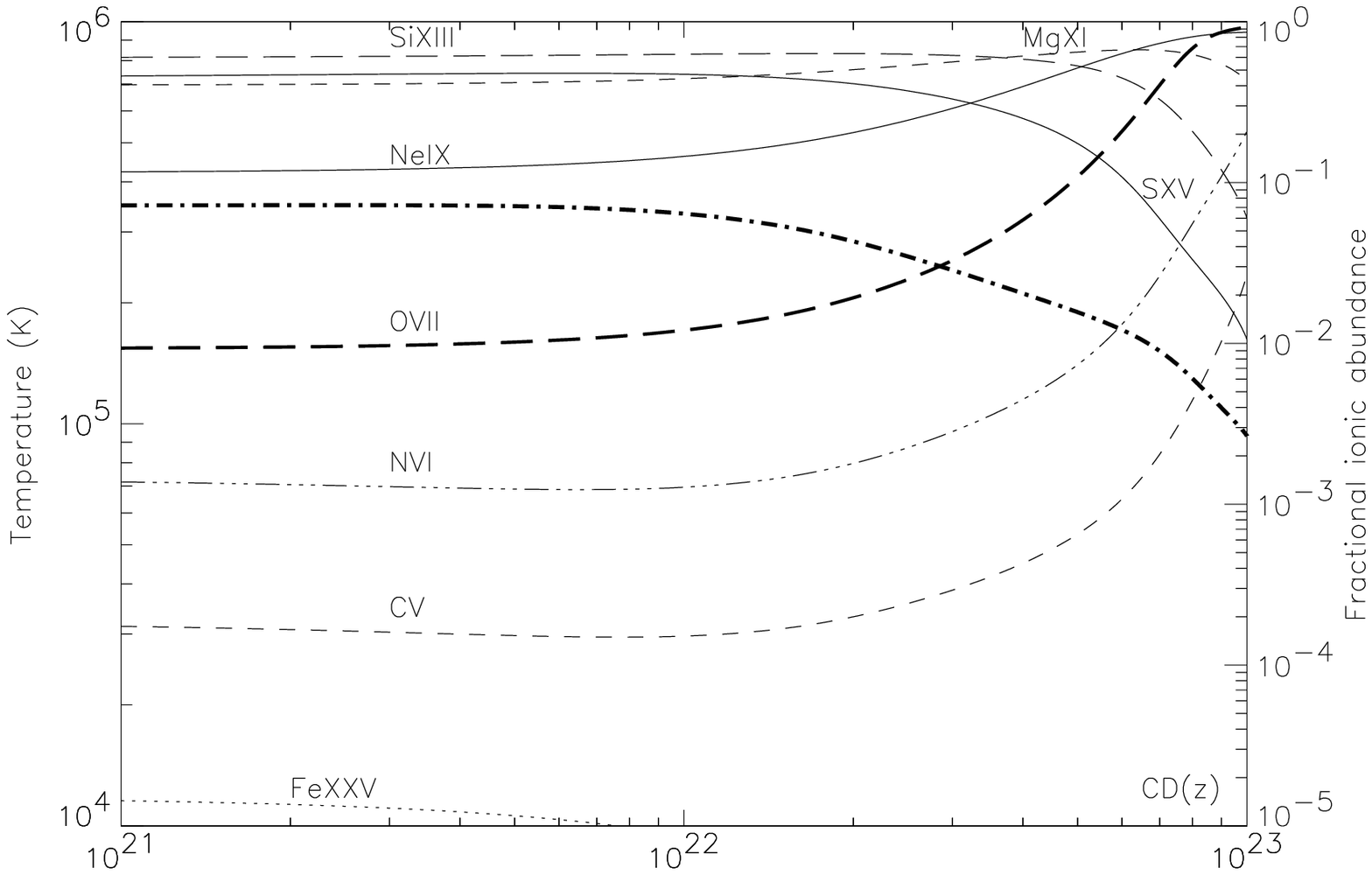,width=7.5cm,height=7.cm}
\vspace{-0.5cm}
\psfig{figure=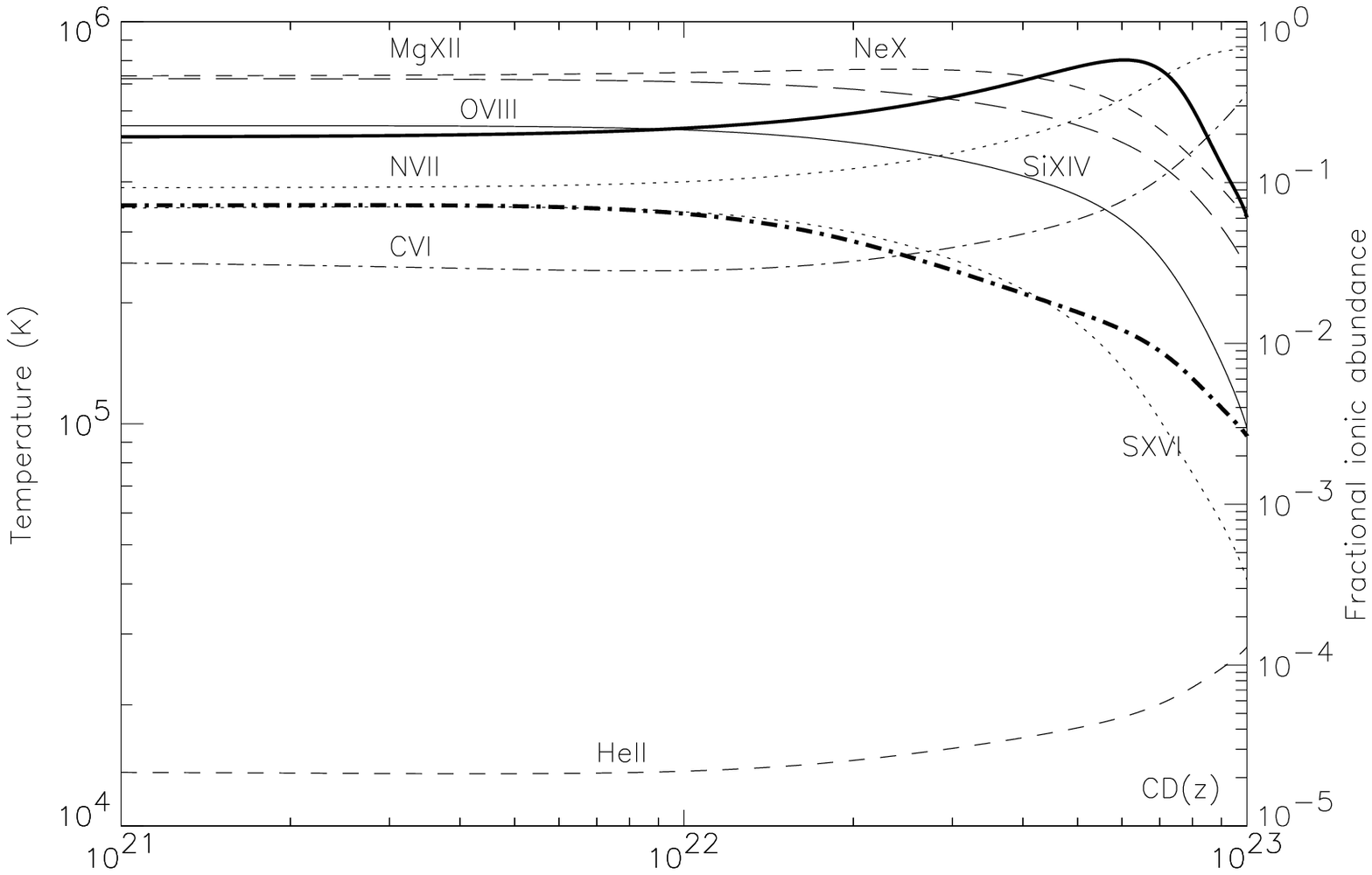,width=7.5cm,height=7.cm}
\caption{Same caption as Fig.\ref{fig1}, for a model with $CD=10
^{23}$ cm$^{-2}$, $n_H=10^{7}$ cm$^{-3}$, and 
$\xi=100$\,(Model\,2).
}
  \label{fig2}
\end{center}
\end{figure}

  To illustrate these effects which are important for the 
following discussion, Figs. \ref{fig1} and \ref{fig2} display the 
temperature and the fractional ion abundances versus the column 
density computed from the surface to the local point, $CD(z)$, for two 
models: Model 1 has a column density of 3 10$^{25}$ cm$^{-2}$, a 
density of 10$^{12}$ cm$^{-3}$, an ionization parameter of 
$\xi=1000$, and is appropriate for the irradiated atmosphere of an 
accretion disc in an AGN;  Model 2 has a column density of 10$^{23}$ 
cm$^{-2}$, a density of 10$^{7}$ cm$^{-3}$, an ionization parameter 
of $\xi=100$, it is appropriate for the Warm Absorber. The figures at 
the top show the fractional abundances of the oxygen ions, the 
central figures show the fractional abundances of the 
He-like ions, and the bottom ones show the fractional 
abundances of the H-like ions.
 
 Let us examine first Fig. \ref{fig1} corresponding to Model 1. 
 The column density of the hot skin is equal to 10$^{24}$ cm$^{-2}$ 
and its temperature is 2 $10^6$K. All the ``light" elements (C, N, O) 
are completely ionized, and heavier elements are in the form of 
H-like (Si, S) or He-like species (Fe). The column density of the 
intermediate zone is equal to 2 10$^{24}$ cm$^{-2}$, and the 
temperature decreases down to 4 $10^4$K. Oxygen is in the form of 
OVIII and OVII, and in particular in the region where OVII dominates, 
the temperature varies from a few  $10^5$K to a few  $10^4$K. 
It is worth noticing that the column density of the OVII region is 
very large in this model (about 10$^{21}$ cm$^{-2}$), but 
the region giving rise to recombination lines (therefore to the OVII 
triplet) is much smaller, as OVII and OVIII do not coincide in 
position. The OVII spectrum is therefore emitted by the intermediate 
zone where the temperature varies strongly.
The rest of the slab is cold and contributes only to the UV emission.

Fig. \ref{fig2} corresponding to Model 2 shows that, owing to the 
relatively low value of the ionization parameter, the hot skin is 
restricted to a column density of about  10$^{22}$  cm$^{-2}$, and 
its temperature is relatively low (3.5 $10^5$K). The intermediate 
zone is not complete and the temperature decreases from 3.5 $10^5$K 
at the surface to $10^5$K at the back. OVIII ions dominate in a 
fraction of this  zone, but OVII is abundant only near the back 
where the  temperature is lower. Again the OVII spectrum is formed in 
this  inhomogeneous zone. There is no cold zone at all.

From these examples one can understand that {\it the emission of a 
thick medium is not easily described by a homogeneous model}. For the 
formation of the He-like triplet, the column density which matters is 
that of the H-like ions, while it is the temperature of the He-like 
zone which determines the collisional excitations of the $2s ^3S_{1}\ 
-\ 2s\ ^3P_{0,1,2}$ transitions, important for the $G$ ratio.  One 
also sees that there will be a ``saturation" of the line fluxes due 
to the fact that the size of the emission regions is limited 
according to the value of the ionization parameter. One can guess 
thus that the same conditions for a given ion (same column density, 
same temperature) could be obtained with different ionization 
parameters and column densities.

\begin{figure}
\begin{center}
\psfig{figure=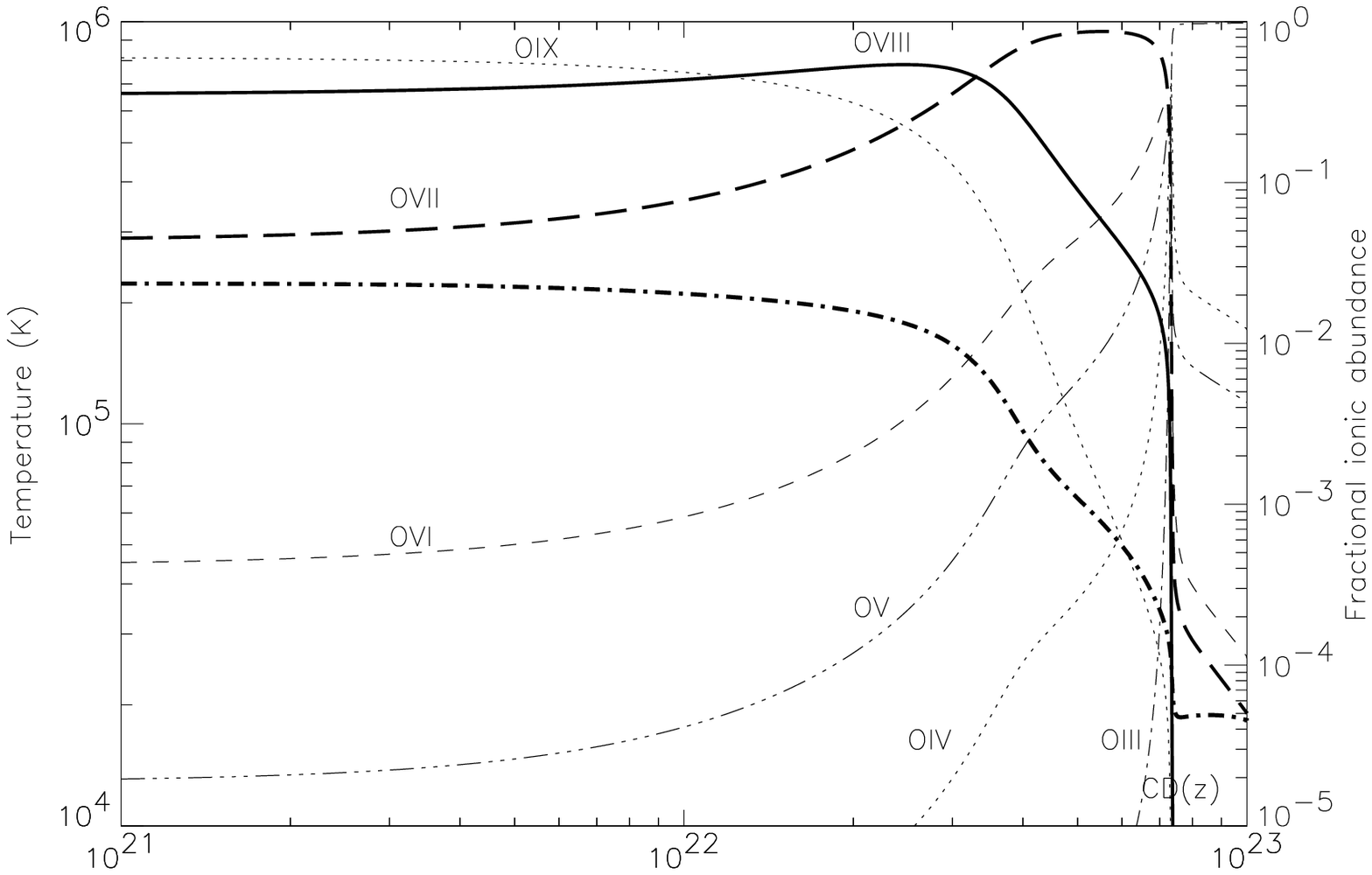,width=7.5cm,height=7.cm}
\vspace{-0.5cm}
\psfig{figure=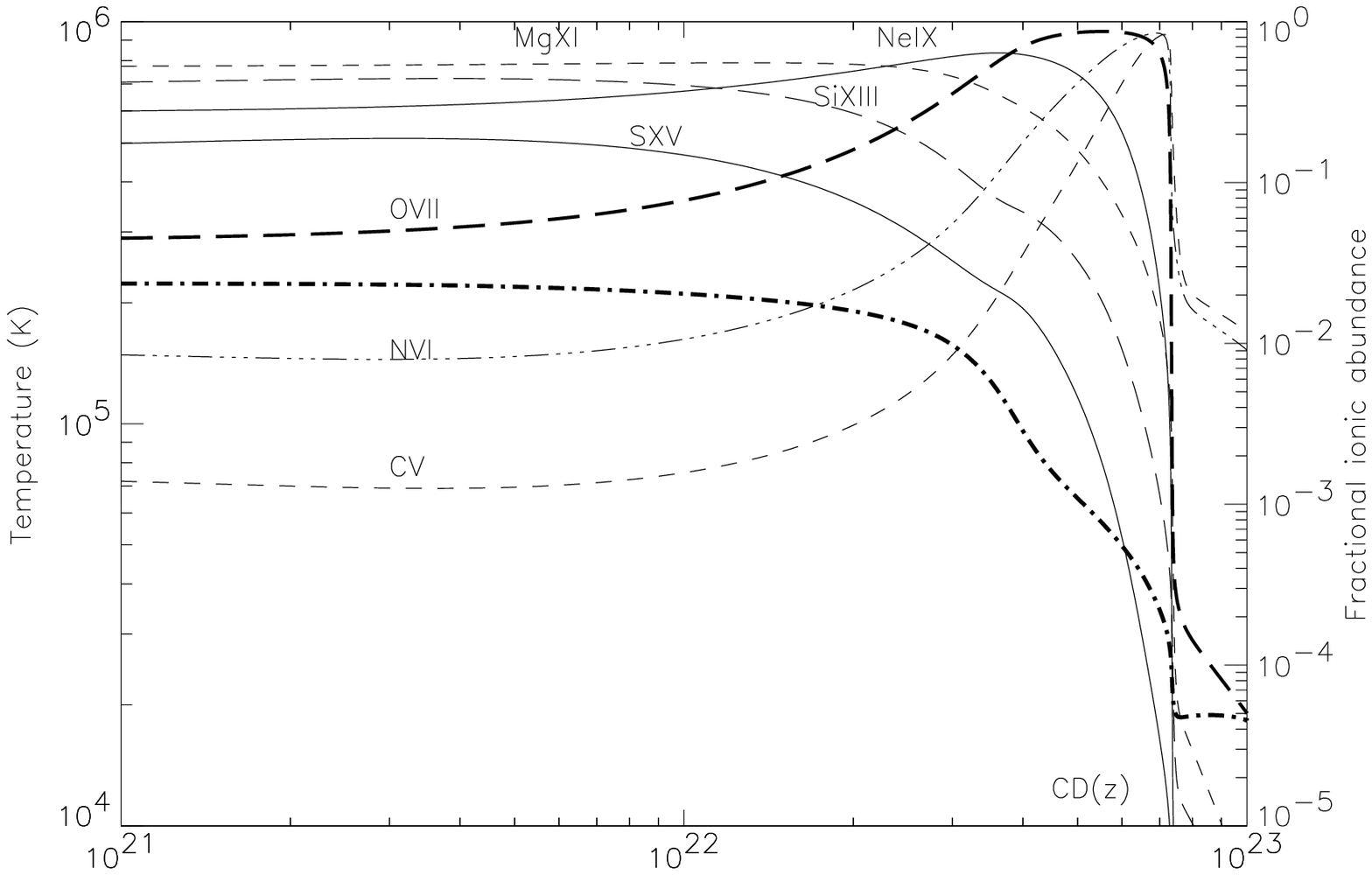,width=7.5cm,height=7.cm}
\vspace{-0.5cm}
\psfig{figure=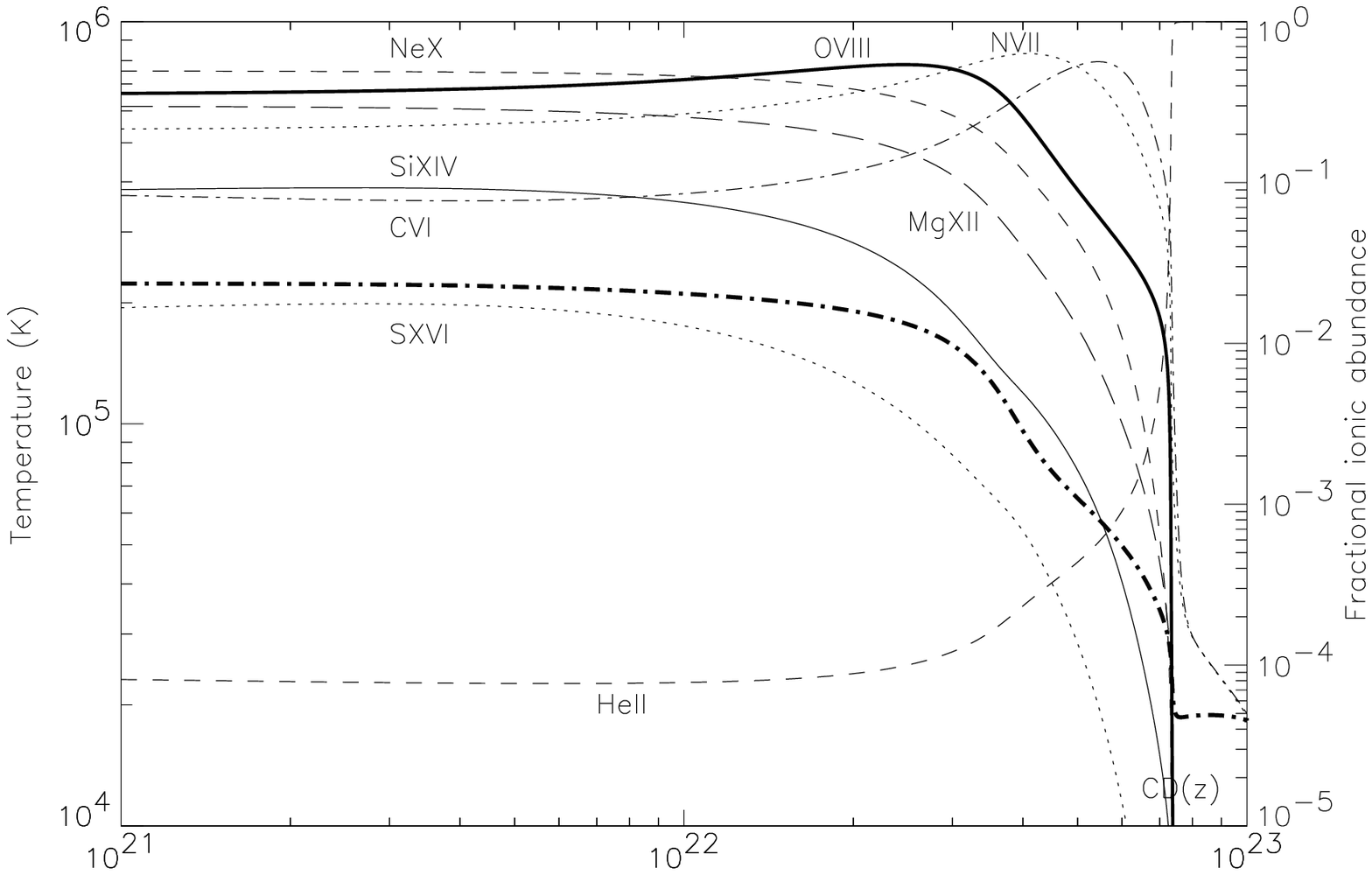,width=7.5cm,height=7.cm}
\caption{Same caption as Fig.\ref{fig1}, for a model with 
$CD=10^{23}$ cm$^{-2}$, a density of $n_H=^{7}$ cm$^{-3}$, 
photoionized by the AGN continuum with
$\xi_{eq}=100$.
}
  \label{fig3}
\end{center}
\end{figure}

The spectral distribution of the incident continuum must also be 
taken into account. To illustrate its influence, Fig. \ref{fig3} 
displays the same curves as Figs. \ref{fig1} and \ref{fig2},  for a 
column density 10$^{23}$ cm$^{-2}$, a density 10$^{7}$ cm$^{-3}$, but 
for the AGN continuum with an equivalent ionization parameter 
$\xi_{\rm eq}=100$.
 They can be  compared to Model 2.
Note also that the temperature is lower in the case of the 
AGN continuum than in the case of the standard continuum.
This will be translated into differences in the line intensities as 
we will see below.

To complete the information, Fig. \ref{fig-colion-CD} displays the 
ion column densities of a few He-like and H-like species, $CD_{\rm ion} $, 
for slabs of different total column densities $CD$ and for 
different values of the ionization parameter. They are independent of 
the density  in the range $n_{H}=$ 10$^{7}$ to  10$^{12}$ cm$^{-3}$  
because collisional processes are negligible for ionization and 
recombination. The long dashed line marks the ion 
column density of 10$^{17}$ cm$^{-2}$.
 Several important conclusions can be drawn from these curves.
 
All ion column densities saturate at large values of $CD$ 
\footnote{This is actually not obvious in the figure for light 
elements and small ionization parameters. Indeed when $\xi$ is small 
and the column density is large, the intermediate zone is dominated 
by weakly ionized species (CII, OII, etc\ldots) and our code Titan 
does not converge correctly, owing to the oversimplified description 
of these ions. This is why the column densities of CV, CVI and OVII, do 
not display the saturation for $\xi=10$ and 30,
 though $CD_{CV}$, $CD_{CVI}$ and $CD_{OVII}$ cannot exceed those 
shown  on the figure.}. As explained above, it is due to the fact 
that the column density of the hot and intermediate zones,  and 
consequently also the column density of the ion species, are 
determined and limited by the value of $\xi$. For $\xi\le 30$ the 
column density of the intermediate zone is small and for instance  
$CD_{OVII}$  saturates at a few $ 10^{19}$ cm$^{-2}$. This result is 
important, as it can give an insight on the ionization parameter. If 
one finds from the observation of emission lines that 
$CD_{OVII}$ is larger than 10$^{20}$ cm$^{-2}$, it means that $\xi$ 
is larger than 300.

Another striking result is the rapid increase of $CD_{CV}$ and 
$CD_{OVII}$ for large values of $CD$ and of $\xi$: they can reach 
values as large as 10$^{21}$ cm$^{-2}$, contrary to heavier  
He-like species. This behaviour can be understood when  considering 
Figs. \ref{fig1}, \ref{fig2}, and  \ref{fig3}. For a given $\xi$, 
the region where a given ion species dominates is located at a 
specific position inside the slab. When the column density is small, 
the ion is not dominant, and $CD_{\rm ion} $ is approximately proportional 
to $CD$. When $CD$ increases, it reaches the position where the ion 
becomes dominant, so $CD_{\rm ion} $ increases more rapidly. 
 $CD_{\rm ion} $ stops increasing when $CD$ becomes larger and the element 
becomes less ionized. 
 
We see also that it is possible to get {\it the same value of  
$CD_{ion} $ for a large range of values of the total column density 
and of the ionization parameter}. For instance $CD_{OVII}=10^{17}$ 
cm$^{-2}$ can be obtained with $CD=10^{21}$ cm$^{-2}$ and $\xi=10$, or 
with $CD=5\ 10^{23}$ cm$^{-2}$ and $\xi=300$.

One can ask whether these results depend or not on the ionizing 
spectrum. 
Fig. \ref{fig-colion-CD} shows a model computed with the AGN 
continuum and an equivalent ionization parameter $\xi_{eq}=100$ (the 
diamond on the figure). We see that the result does not differ 
strongly from that obtained with the standard continuum, but of 
course a more extensive study is required to be obtain a firm 
conclusion. 
If these curves are indeed almost independent of the ionizing 
spectrum, they would help to determine the ionization parameter in 
the Warm Absorber, where the column densities of the ion species are 
measured through a curve-of-growth analysis.
   
   For each ion, the temperature can be measured directly through the 
shape of the corresponding radiative recombination continuum (RRC), 
which is expected to vary with the frequency as $exp(-h\nu/kT)$ (cf. 
Kinkhabwala et al. 2002). Note again that there is some 
ambiguity here for an inhomogeneous medium.
 First the measurement concerns the H-like zone, while the relevant 
temperature is that of the He-like zone. Second, in the case of very 
thick models (the accretion disc atmosphere for instance)
 the shape of the RRC is no 
longer a pure exponential, owing to photon reabsorption. 

It is interesting to compare these observations with the temperature 
of the He-like and of the H-like zones.  As an example, Fig. 
\ref{fig5} displays the maximum temperatures  $T_{max}$ of the 
regions where OVII and OVIII are dominant, versus the OVII column 
density $CD_{OVII}$, for models of various column densities,
 including a few with a 
non-standard incident continuum. The solid lines correspond to 
constant values of $\xi$ (thin line for OVIII, thick dashed line for 
OVII), and the thin dot-dashed lines to constant values of $CD$. The 
results are again almost independent of the density  in the range 
$n_{H}=$ 10$^{7}$ to  10$^{12}$ cm$^{-3}$ because the energy losses 
are dominated by recombination lines and continua for which 
collisional processes are negligible. 
Note that the two points in the middle of the figure correspond to 
the AGN continuum, for $CD=10^{23}$ cm$^{-2}$ and $\xi_{eq}$ = 
200. They are located close to the values found with the standard 
continuum for the same parameters, owing to the strong connection 
between the ionization and the thermal equilibrium. Thus {\it the 
$T/CD$(OVII) diagram is almost model-independent}.    

Several conclusions can be drawn from this figure. For high 
values of $CD_{OVII}$, $T_{max}$(OVII) and  $T_{max}$(OVIII) differ 
by a large factor. For a given total column density, $T_{max}$(OVII) 
decreases regularly with increasing ion column density, and for a given 
ion column density, it decreases with $\xi$. $T_{max}$(OVIII) and 
$T_{max}$(OVII) are of 
the order of 10$^{6}$K only for high values of $\xi$ and  $CD_{\rm ion} $ 
smaller than 10$^{18}$ cm$^{-2}$. $T_{max}$(OVII) is smaller than 10$^{5}$K for 
small values of $\xi$, and consequently for small values of the ion 
column density, ($\le10^{18}$ cm$^{-2}$). This behaviour can be 
understood with the help of Figs. \ref{fig1} and \ref{fig2}. 
 
\section{Results: the spectrum} 

\subsection{Total line intensities}

It is important to know which models may lead to detectable line 
emissions. In this aim we have computed the equivalent widths, EWs, 
of the {\it sum} of the triplet He-lines {\it with respect to the 
incident continuum}. It is a good measure of the EWs when the 
continuum is seen directly because the contribution of the slab 
itself is negligible except for very thick slabs\footnote{These EWs 
correspond to a non-radial view of the objects.}. It is obviously an 
underestimation in the case of Seyfert 2 with a hidden source of 
continuum. For instance these EWs should be multiplied by a factor 
100 if the hidden continuum is reflected by a medium with a Thomson 
thickness of 0.01 like in NGC 1068. A line can be detected when its 
EW is larger than 0.1 eV.

 Figs. \ref{fig6} and \ref{fig7} display these EWs for all the 
He-like species and for several values of the ionization parameter 
as a function of the column density of the slab, for the standard 
continuum. Again they are almost independent of the density (because 
we consider the sum of the triplet terms and not the different 
components). As expected, the light species with the lowest 
ionization potential (CV, NVI, OVII) are the most intense for the 
smallest $\xi$, and on the contrary FeXXV is the most intense line 
for the largest $\xi$ and it disappears for $\xi \le 100$. 
 
All EWs saturate at large column densities. This is easy to 
understand. These lines are due to recombination, and they saturate 
because the number of recombinations does not increase above a given 
column density, if the H-like ions are completely recombined. 

 For clarity, only one result for the AGN continuum is shown on Fig. 
\ref{fig7}, but it is enough to see that the EWs of this model are 
quite different from those of the standard continuum. 
So {\it these curves cannot be considered as universal ones, contrary 
to $T/CD$(OVII) diagram}.

 Fig. \ref{fig8} recapitulates the EWs of the OVII triplet versus the 
column density for all models including a few with the AGN 
continuum. The saturation effect is visible for all ionization 
parameters, and we see that the EW of the OVII triplet cannot exceed 
25 eV for the standard continuum (this also seems to be the case 
for the AGN continuum, but we need more models to confirm this). 
This is an important result, as it indicates that a larger observed OVII 
would imply a different kind of model (for instance, non-stationary), 
or a partial extinction of the continuum.
 Note that the same EW can be obtained with a small $CD$ and a small 
$\xi$, or with a large $CD$ and a large $\xi$. It is thus obvious 
that {\it from the measure of the CV-NVI-OVII triplets only, it is 
not possible to determine the values of the column density and of the 
ionization parameter}, even if the temperature of the 
emitting region can be measured. A detailed study of the whole 
spectrum, including lines of heavier elements, is clearly necessary 
to get the physical parameters of the emissive medium. 

Finally the EW of the OVII triplet for the AGN continuum is also 
shown on Fig. \ref{fig8}. They are quite different from 
the standard continuum. This is due to the different values of the 
He-like over H-like abundance ratios in the emission regions (cf. 
Fig. \ref{fig3}).

\subsection{The G and R ratios}

Fig. \ref{G-ratio} displays the $G$ ratio for CV, OVII, SiXIII, and FeXXV
 versus the column density $CD$,
for different values of the ionization parameter,
and for the standard continuum. One model
corresponding to the AGN continuum is also shown, for $\xi_{eq}=100$,
and we see that it fits relatively well the curves obtained for the
standard continuum. 

This figure shows the strong dependency of the
G-ratio on the column density, for a given ionization parameter.
This is because the resonance line $w$ decreases with $CD$ 
(and therefore $G$ increases),
 due to the decrease of the incident radiative excitation.
For a very large column density, the value of $G$ reaches a constant value
due to the limit of the ion column density (see Fig.4).

Note also that at large values of the column density, another process occurs, namely
photon destruction, which is more important 
for resonance lines than for forbidden lines due to resonant scattering.
 This destruction is mainly
due to photoionization of ions of lower ionization energy by the
line photons. Correct values of the ratio cannot be computed without
taking into account these interactions with other ions.
 It is not properly taken
into account by escape probability approximations (cf. Dumont et al.
2003).

The figure also shows that the dependency of $G$ on the ionization 
parameter. This behaviour is easy to explain. If we consider light
elements, we know (Fig. 4) that for a given column density,
$CD$(CV) and $CD$(OVII) decrease with increasing $\xi$ until they reach 
saturation, because the emission line region is pushed towards the back of the 
slab. Consequently radiative excitations of the resonance line 
increase, and the $G$ ratio decreases. On the contrary, $CD$(FeXXV) 
increases with $\xi$, inducing the decrease of $G$. 

An important point is worth noticing.  {\it The $G$ ratio is 
almost never equal to the recombination value} ($\sim 4.5$ for OVII for 
instance). 
 At high column densities, we have seen that $G$ is larger than the 
recombination value, owing to the influence of the destruction 
mechanism on resonance lines. At small column densities, 
 $G$ is much smaller than the recombination 
 value (owing to the 
influence of radiative excitations on the $w$ line, as explained above). 
 $G$ tends to a constant value at small 
values of $CD$, because the intensities of $z$, $x$ and $y$ are proportional 
to the rate of photoionization, while $w$ is proportional to the rate 
of radiative excitation, both being proportional to the incident 
flux, when it is not absorbed. 
We see that this value is not reached at  $CD=10^{18}$ cm$^{-2}$  
for low values of $\xi$,
 because 
the optical thickness at the line center is still higher than unity, 
so the rate of radiative excitations is smaller than at 
the surface.

Actually, an object like NGC 1068 where the 
$G$(OVII)
ratio is of the order of 7 (cf. Kinkhabwala et al. 2002)
can be explained as well with $\xi\sim 10$ and $CD\sim  10^{21}$ 
cm$^{-2}$, as with $\xi\sim 300$ and $CD\sim  10^{24}$ 
cm$^{-2}$. The consideration of other spectral features due to other 
ions is required
 to disantangle these possibilities.

 Fig. \ref{fig-R-OVII-SiXIII}
displays the $R$ ratios for OVII and SiXIII as a function of the
ionization parameter, for two values of the column density, and for a
density $n_H=10^{7}$ cm$^{-3}$. As expected, the $R$ ratios are almost
 independent of the column density
and of the ionization parameter, as they are not affected by radiative 
excitations.  Fig. \ref{R-ratio} also displays the $R$ ratios of
 OVII and FeXXV as a
function of the density, for two values of the column density and of the
ionization parameter, for the standard continuum. It confirms
the well-known
result that $R$ decreases with an increasing density, as a consequence of
collisional excitations of the $2s\ ^3S_1\ - 2p\ ^3P$
transition. Note that the effect does not appear in the FeXXV diagram, 
as the critical density for this ion is larger than our limit 
$n_H=10^{12}$ cm$^{-3}$.

Finally, all this discussion is illustrated in Figs. \ref{Spectre_raieO7_n7} and
\ref{Spectre_raieSi13_n7} which show the spectra of
the OVII and SiXIII triplets for several values of the column density
and of the ionization parameter. The lines are given for a spectral
resolution of $\frac{\mathrm{E}}{\Delta \mathrm{E}}=500$. The figures
show the influence of an increasing column density at a given ionization
parameter (left panels), and the influence of an increasing ionization
parameter at a given column density (right panels).

\subsection{The whole spectrum}

As mentioned in the introduction, several types of objects are now 
observed with a great spectral resolution from 200 eV to 1 keV by 
Chandra and XMM-Newton. For instance, a pure reflected spectrum is 
observed in Seyfert 2 galaxies. It is thus interesting to give 
the reflected spectrum in this energy range as an 
illustration, for a few 
values of the column density and of the ionization parameter. 
A grid of such spectra will be used in a future paper in a study of 
NGC 1068 and Mrk 3. 

These spectra are displayed in Fig. \ref{spectres}, convolved with a 
gaussian profile having a FWHM equal to 1500 km/sec. Note that the 
fluxes are given in $\nu F_{\nu}$, which is equivalent to 
$F_{\lambda}$ in photon number per unit of $\lambda$. These spectra 
can thus be directly compared with the published spectra (see Kimkhabala 
et al. 2002 for instance). 

The three figures at the top display the spectra of models computed 
with the 
AGN continuum, for a given large value of the column density ($10^{23}$ 
cm$^{-2}$), and for large values of the ionization parameter. They show
 for instance that the relative intensities of the lines from
 different ions vary with the ionization parameter, as well as the relative 
 intensities of the triplet lines from a given He-like ion. The OVIII 
 L$\alpha$ line is the most intense.
 
 The two 
 figures at the bottom display the spectra for models computed with 
 the standard 
 continuum, for low values of the column density and of the 
 ionization parameter (respectively $CD=10^{22}$ 
cm$^{-2}$ and $\xi=30$, $CD=10^{21}$ 
cm$^{-2}$ and $\xi=10$). The spectra are dominated by the OVII 
triplet, whose relative intensities are the same for both cases, while 
the relative intensities of the other features vary (in particular 
note the strong CV RRC and the very weak OVIII L$\alpha$ line in the 
bottom spectrum). 

If there are 
 several regions with different values of $\xi$ and of $CD$
  contributing to the spectrum, it is thus necessary to sum the {\it 
  complete}
  spectra emitted by each region, to fit the whole observed spectrum.
   {\it It is in particular clear that  
 the different ions cannot be considered independently, as each emission 
 region provides lines from several ions}.

\section{Conclusion}

Using our new version of the photoionization code Titan which is 
unique in treating 
the full line transfer, we have computed the line 
intensities emitted by an X-ray photoionized plasma.  In particular, 
it was shown in Dumont et al. 
(2003) that much smaller values of the X-ray line intensities are 
obtained for resonant lines than with the other photoionization codes 
where the line transfer is replaced by the escape probability 
approximation. We focused on the 
He-like emission, for a range of 
ionization parameters and column densities. The  
results of this study are:

\begin{itemize}

\item The column density of a given ion species increases with the 
total column density of the medium until it reaches a limiting value 
depending on the ionization parameter.

\item In the range of ionization parameters considered here ($\xi$ 
varying from 5 to 1000), 
the column density of light He-like ions (CV-NVI-OVII)
 decreases with increasing ionization 
parameter for a given value of the column density,
 while the opposite is the case for FeXXV.

\item It is difficult to define a unique temperature for the line 
emitting region. Indeed the temperature varies 
accross the region emitting He-like lines, and it differs from that of 
the H-like region, for ion column densities larger than a few $10^{17}$ cm$^{-2}$.

\item For a given ionization parameter, the maximum temperature of the OVII 
emitting region decreases with the OVII column density when it is 
larger than a few $10^{17}$ cm$^{-2}$,  and the relation seems to be 
model-independent. So this is a rough way to determine the ionization 
parameter in the Warm Absorber of AGN, if $T$(OVII) and $CD$(OVII) are 
known. 

\item The equivalent widths (computed assuming that the continuum is 
seen directly, when this is not the case they have to be divided by the 
Thomson thickness of the reflecting medium) of the sum of the He-like triplet increase 
with column density until they reach a limiting value
depending on the ionization parameter. We find for instance that the highest value 
reached by EW(OVII triplet) is $\sim$ 25eV. A larger observed OVII 
would imply different  models (for instance, non-stationary), 
or a partial extinction of the continuum. 
 
\item The $G$ ratios of the He-triplets are almost never 
equal to
the pure
recombination value. For low values of the column density, the ratio is 
smaller, owing to radiative excitation of 
the resonance line, and for high values of the column density  it is 
larger, owing to photon 
destruction during the process of resonant scattering. 

\item The same EWs and the same $G$ ratios can be 
obtained with a small ionization parameter and a small column 
density, or with a large ionization parameter and a large column 
density.  For instance the observed
$G$(OVII)
ratio in NGC 1068
can be explained with $\xi\sim 10$ and $CD\sim  10^{21}$ 
cm$^{-2}$, or with $\xi\sim 300$ and $CD\sim  10^{23}$ 
cm$^{-2}$.

\end{itemize}

To summarize, our study has shown that in modelling X-ray 
spectra one cannot dispense with a treatment including 
all ion species, a full transfer of the continuum and, in the case 
of a relatively thick medium, also of the lines. This is true even
 if there are 
 several regions with different physical parameters 
  contributing to the spectrum (as for instance in the case of an 
  extended diluted medium with a variation of the dilution 
  factor of the incident flux). In particular the different ions 
  cannot be considered 
  independently, as each emission 
 region provides lines from several other ions.

  The present paper presents only the qualitative behaviour of the 
 spectra, as the errors due to the use of the simplified atomic model 
 for He-like ions can reach 50$\%$ (which, incidentally, is smaller 
 than the errors obtained when using the escape probability 
 approximation for a column density of $10^{22}$ cm$^{-2}$ or larger, cf. Dumont et al. 2003, and in 
 preparation). Moreover, we have not discussed in detail 
 the influence of the spectral 
distribution of the ionizing continuum, which is important, nor that 
of turbulent velocity, and the was restricted to the study of a 
constant-density medium. Forthcoming papers will be devoted to building 
a grid of X-ray spectra taking into account these various possibilities, 
using a more sophisticated He-like model-atom (restricted here to 11 
levels plus a continuum), and 
to apply the results to a few well observed X-ray emitting sources. 

  \begin{figure*}
\begin{center}
\psfig{figure=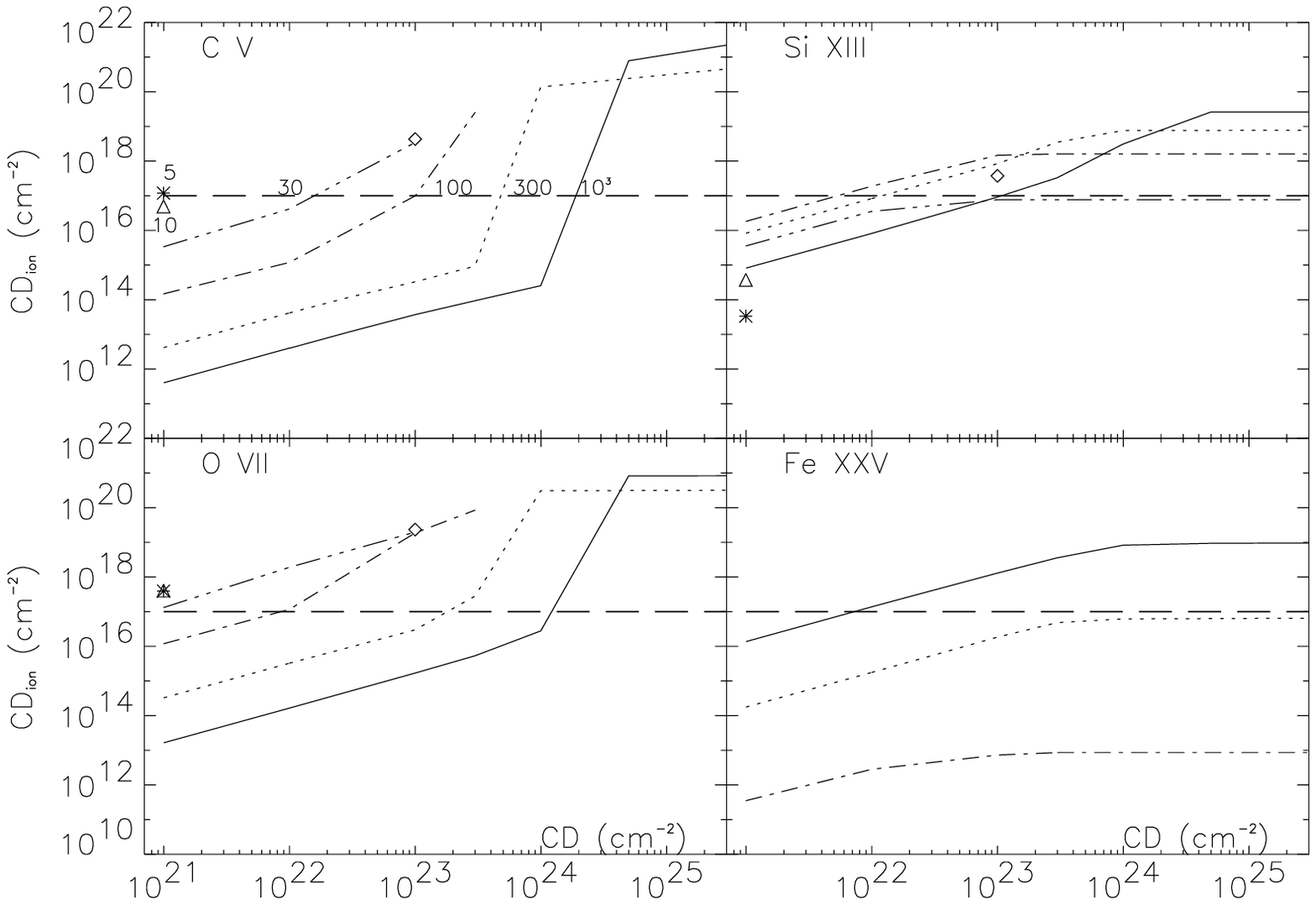,width=18cm}
\psfig{figure=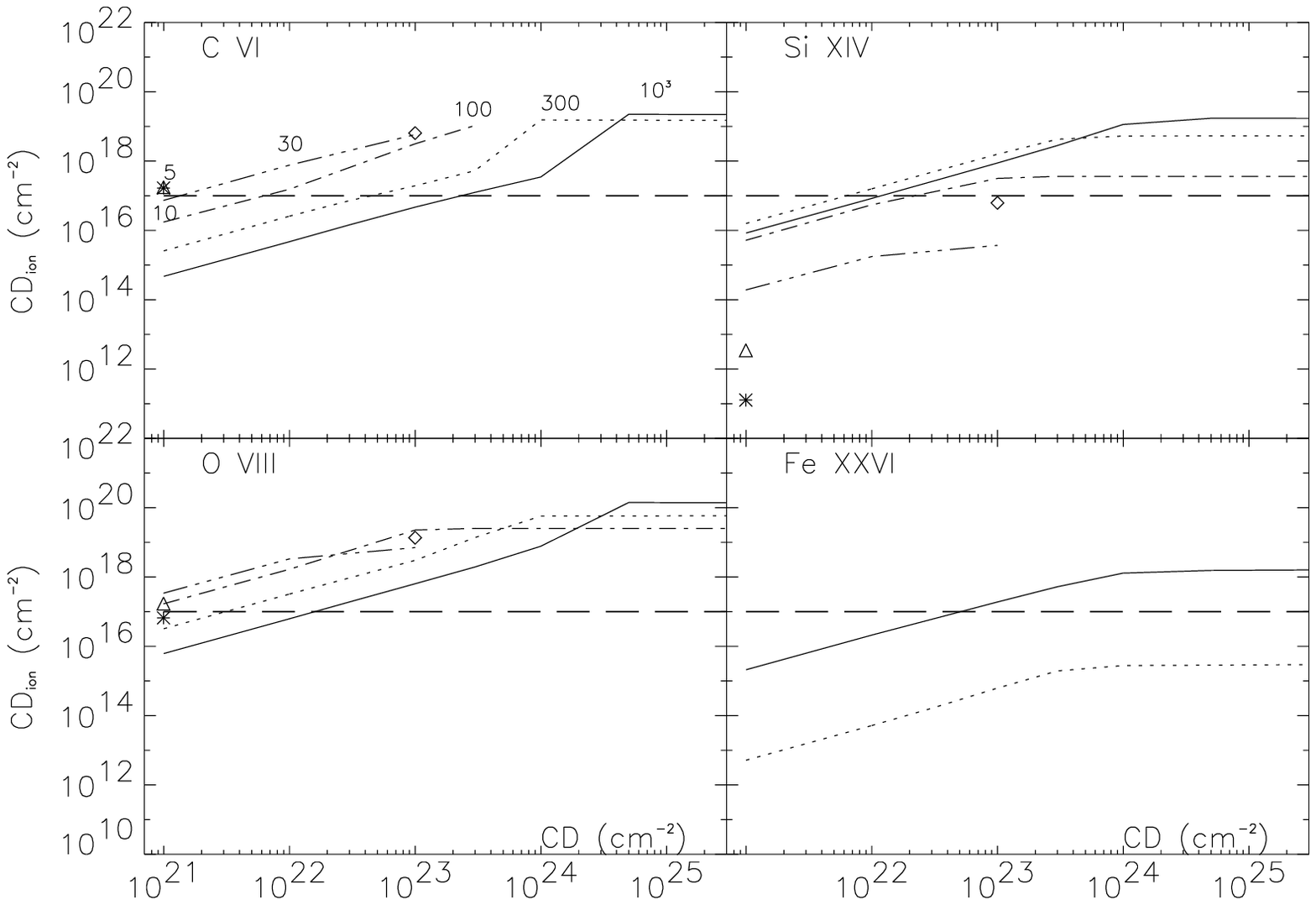,width=18cm}
\caption{Column densities of a few He-like and H-like species
 versus the 
total column density, for different values of $\xi$, for the standard 
continuum. 
These curves are independent 
of the density (from 10$^{7}$ to 10$^{12}$ cm$^{-3}$). The 
diamond corresponds to a slab photoionized by the AGN 
continuum with 
$\xi_{eq}=100$. The long dashed line marks the column density of 
10$^{17}$ cm$^{-2}$.  }
  \label{fig-colion-CD}
\end{center}
\end{figure*}

\begin{figure}
\begin{center}
\psfig{figure=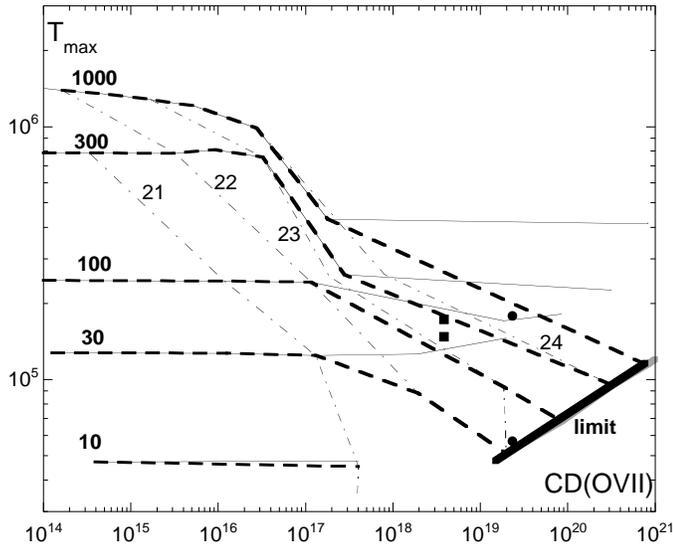,width=9cm}
\caption{Maximum temperatures of the regions where OVII and OVIII are 
dominant, as a
function of the OVII column density in cm$^{-2}$, for all models 
(these values almost do not depend of densities from 10$^{7}$ to 
10$^{12}$ 
cm$^{-3}$). Solid lines: $T_{max}$(OVIII). Thick dashed lines: 
$T_{max}$(OVII). The labels 
on the curves give the value of $\xi$. Thin dot-dashed lines: 
$T_{max}$(OVII) for a given value of
of 
$CD$ (indicated in logarithms on the curves). Isolated points correspond to the 
AGN 
continuum for $CD=10^{23}$ cm$^{-2}$: squares: $\xi_{eq}=200$; 
circles: $\xi_{eq}=100$. The limit of CD(OVII)/$T_{max}$(OVII) is 
given by the very thick line.}
\label{fig5}
\end{center}
\end{figure}

\begin{figure}
\begin{center}
\psfig{figure=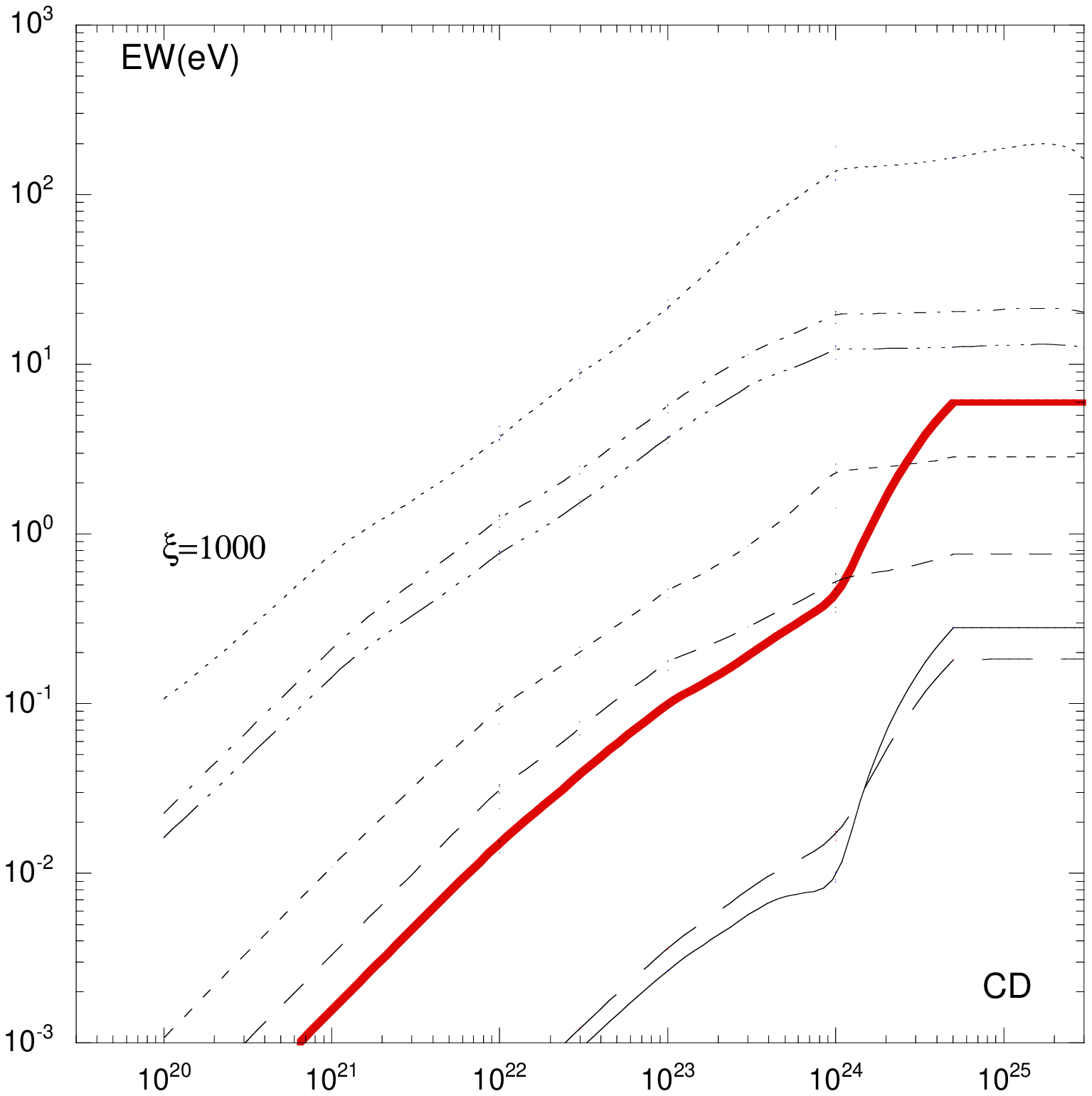,width=7.5cm,height=7.cm}
\psfig{figure=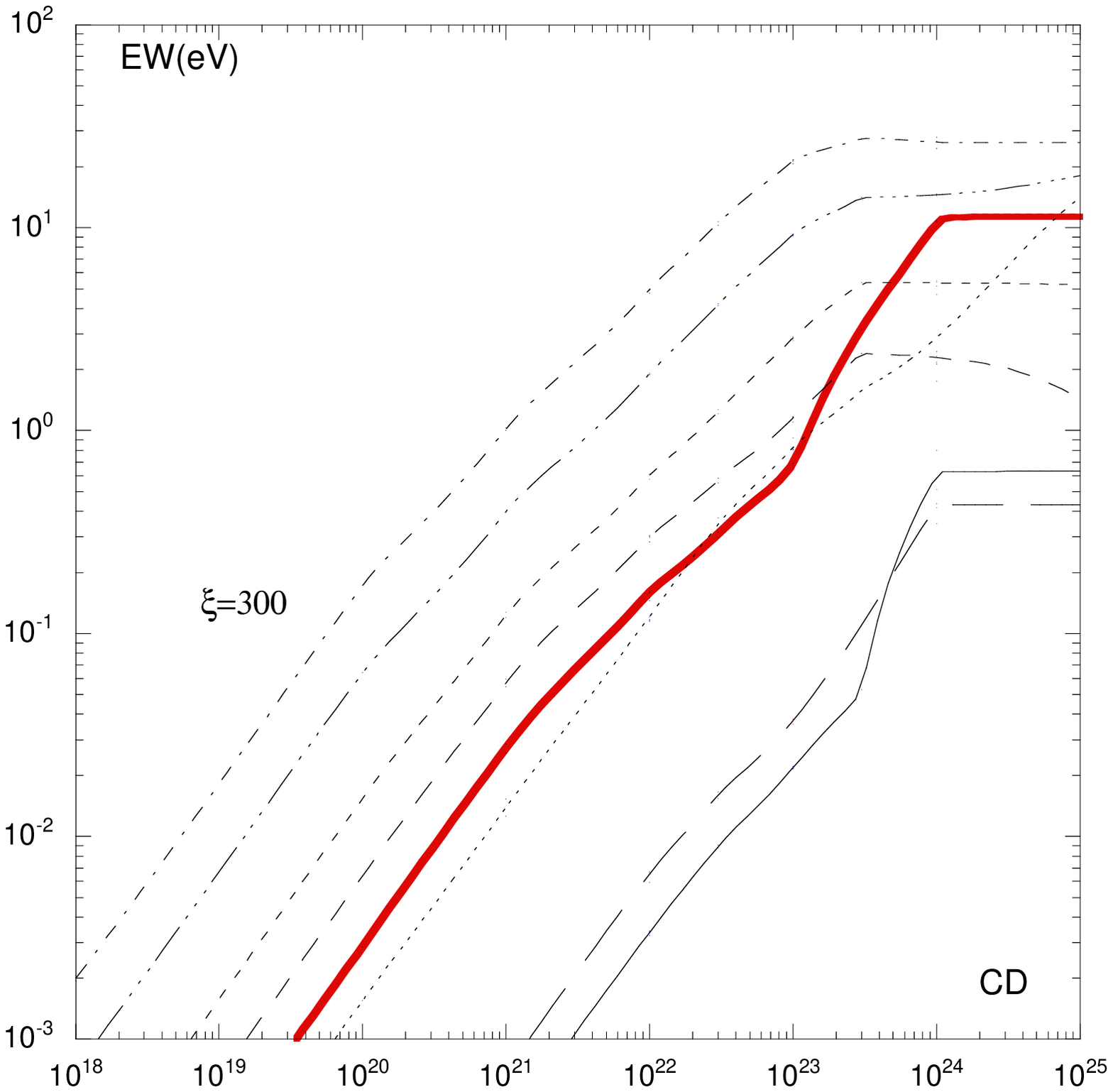,width=7.8cm,height=7.0cm}
\caption{EWs of the sum $w+x+y+z$ for all the He-like species and 
 for $\xi=1000$ and $300$, versus the column density of the slab, for 
the 
 standard continuum, when it is seen directly. 
 The EWs are independent of the density; solid line:C; large dashes: 
 N; thick solid line: O; small dashes: Ne; very small dashes: Mg; 
 dashes and dots: Si; dashes and 3 dots: S; dots: Fe.}
\label{fig6}
\end{center}
\end{figure}

\begin{figure}
\begin{center}
\psfig{figure=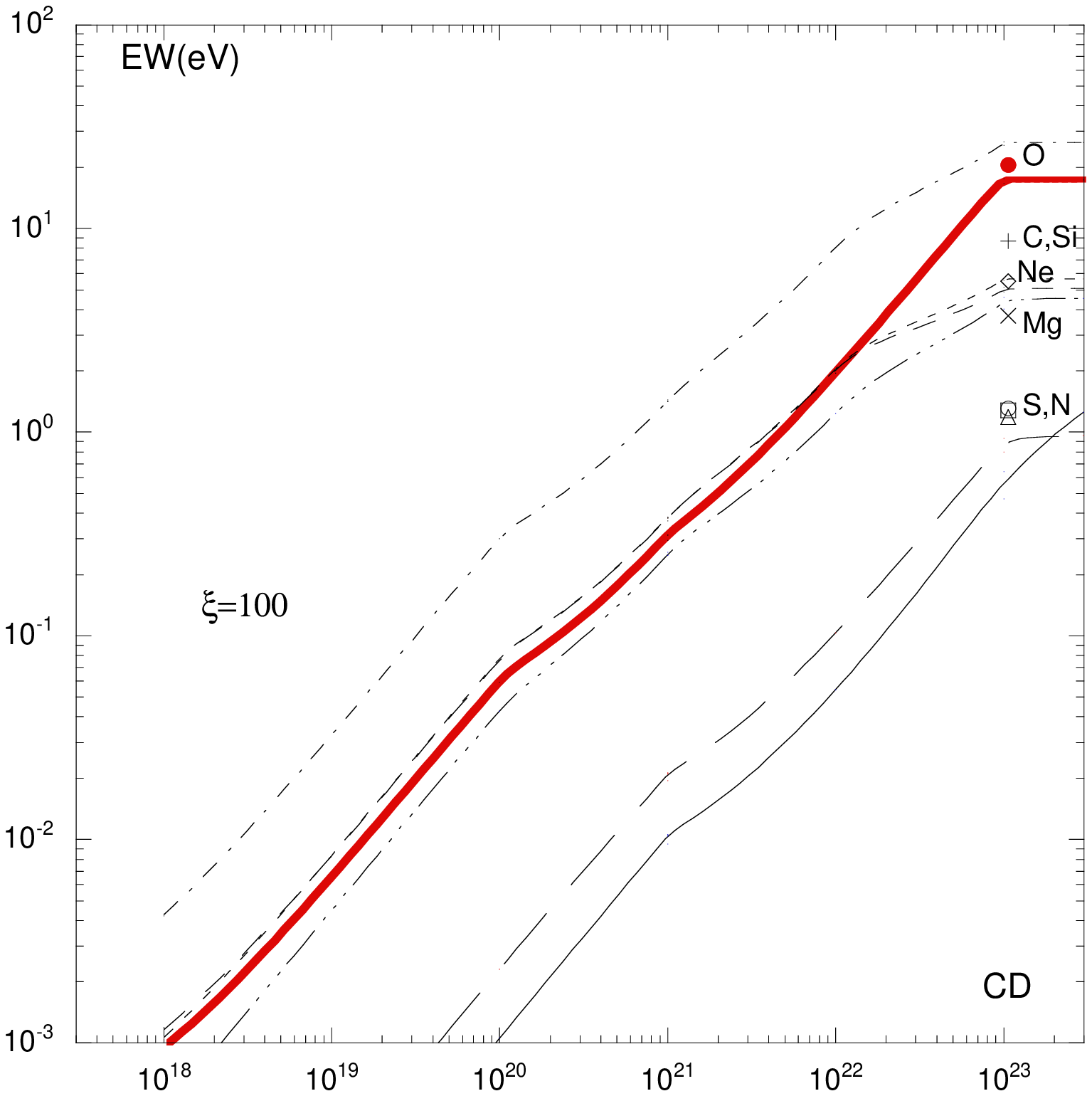,width=8.6cm,height=9.2cm}
\vspace{-2.2cm}
\psfig{figure=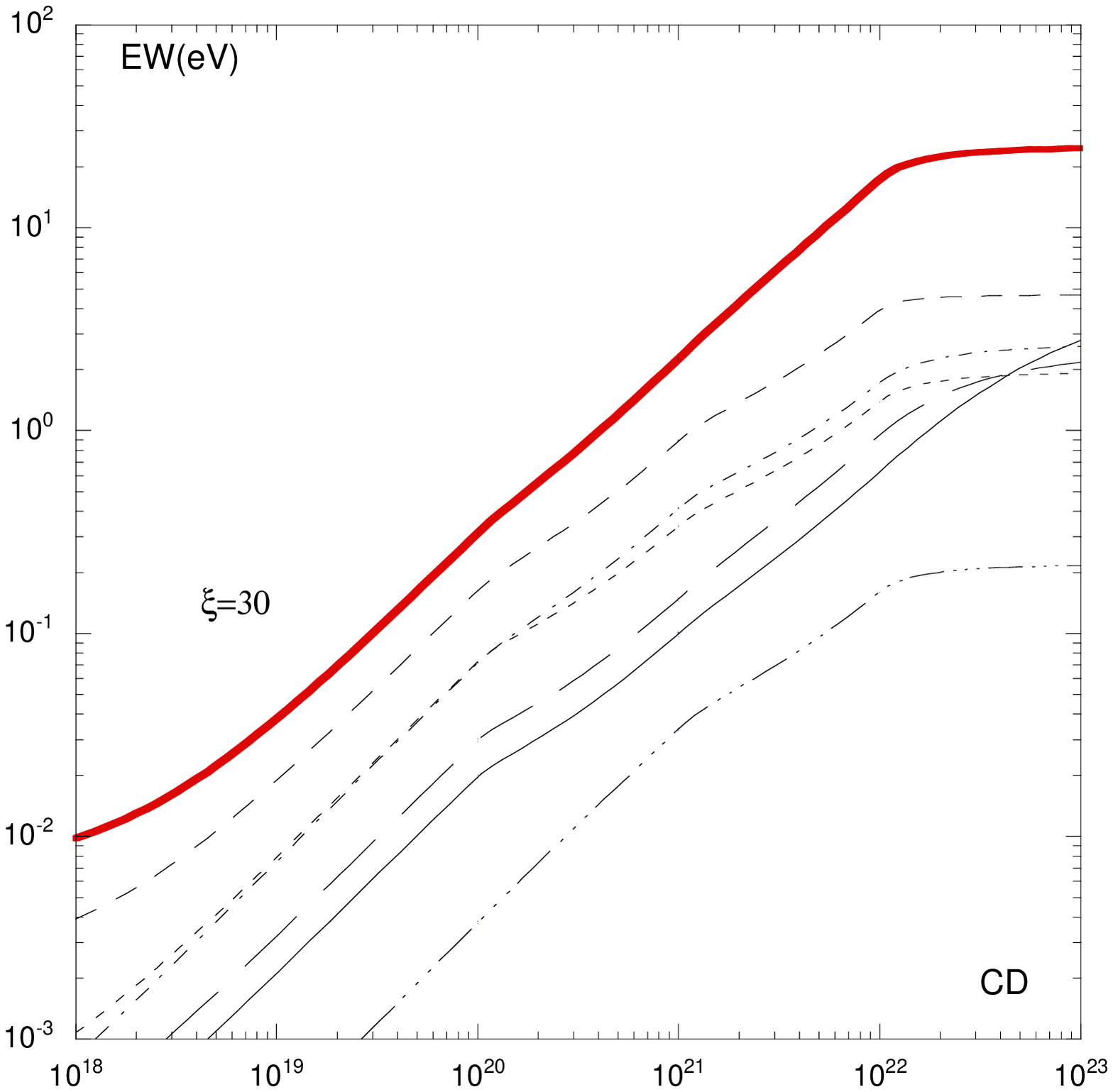,width=8cm,height=7.0cm}
\caption{Same caption as \ref{fig6}, but for  $\xi=100$ and $30$. 
Symbols refer to the AGN model with $\xi_{eq}=100$.}
\label{fig7}
\end{center}
\end{figure}

\begin{figure}
\begin{center}
\psfig{figure=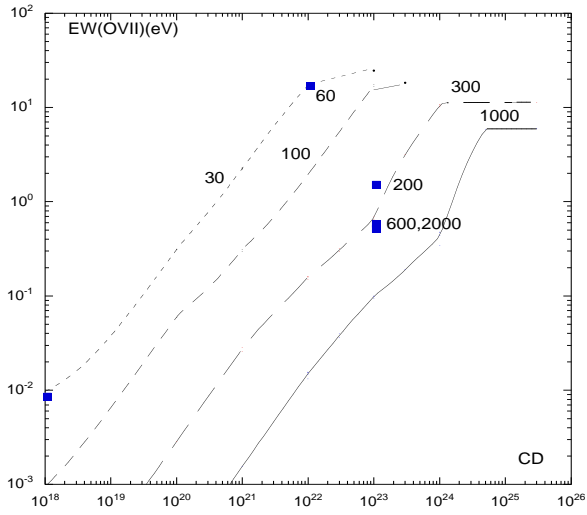,width=9cm,height=9.1cm}
\vspace{-2.1cm}
\caption{EWs of the sum of the OVII triplet for several values of the 
 ionization parameter, including all densities from 10$^7$ to 
 10$^{12}$ 
 cm$^{-3}$, versus the column density of the slab. The curves are 
 labelled with the value of $\xi$. The squares correspond to 
 the AGN spectrum, with $\xi_{eq}$ 
indicated. }
\label{fig8}
\end{center}
\end{figure}

    \begin{figure*}
\begin{center}
\psfig{figure=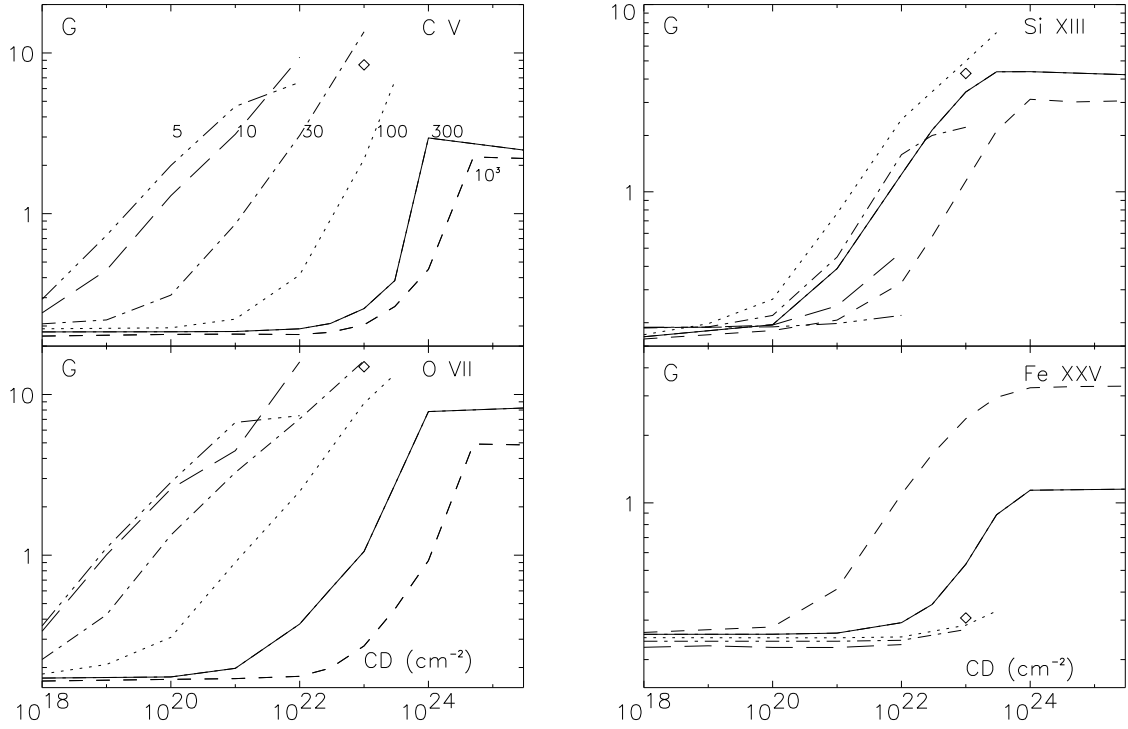,width=16.cm}
\caption{G ratio as a function of the column 
density $CD$ for different values of the ionization parameter $\xi$, 
and for the 
standard continuum. The diamond corresponds to the AGN continuum, for 
$\xi_{eq}=100$.}
  \label{G-ratio}
\end{center}
\end{figure*}

\begin{figure}
\begin{center}
\psfig{figure=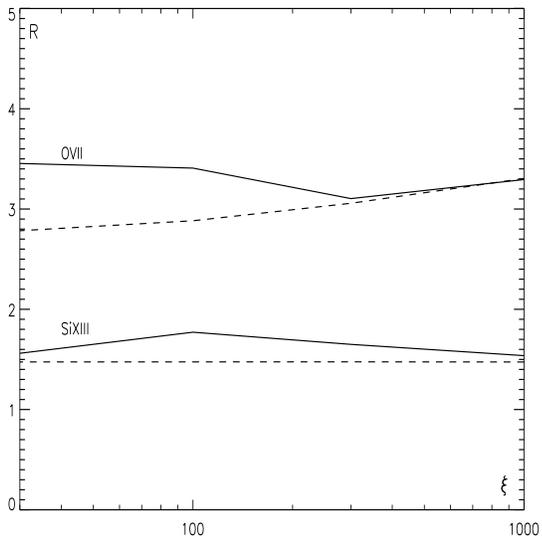,width=8.cm,height=8cm}
\caption{R ratios for OVII and SiXIII as a function
 of the ionization parameter, for a density of 10$^{7}$ cm$^{-3}$, and
for the standard continuum. Solid curves: $CD=10^{23}$ cm$^{-2}$; 
dashed curves:  $CD=10^{18}$ cm$^{-2}$.}
  \label{fig-R-OVII-SiXIII}
\end{center}
\end{figure}

     \begin{figure}
\begin{center}
\psfig{figure=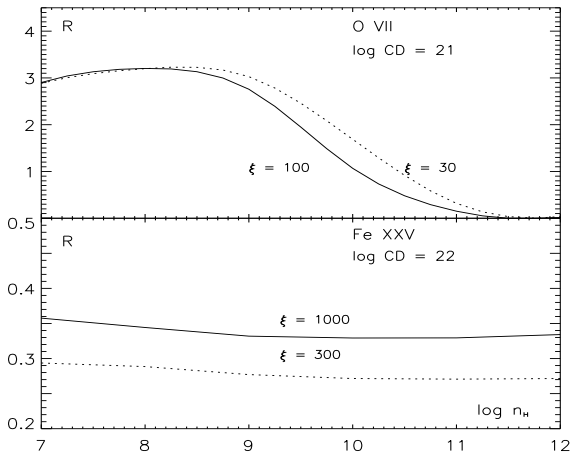,width=11.5cm,height=7.cm}
\caption{R ratios for OVII and FeXXV as a function
 of the density, 
for two values of the column density and of the ionization parameter, 
for the standard continuum.}
  \label{R-ratio}
\end{center}
\end{figure}

\begin{figure*}
\begin{center}
\hspace{0.5cm}\psfig{figure=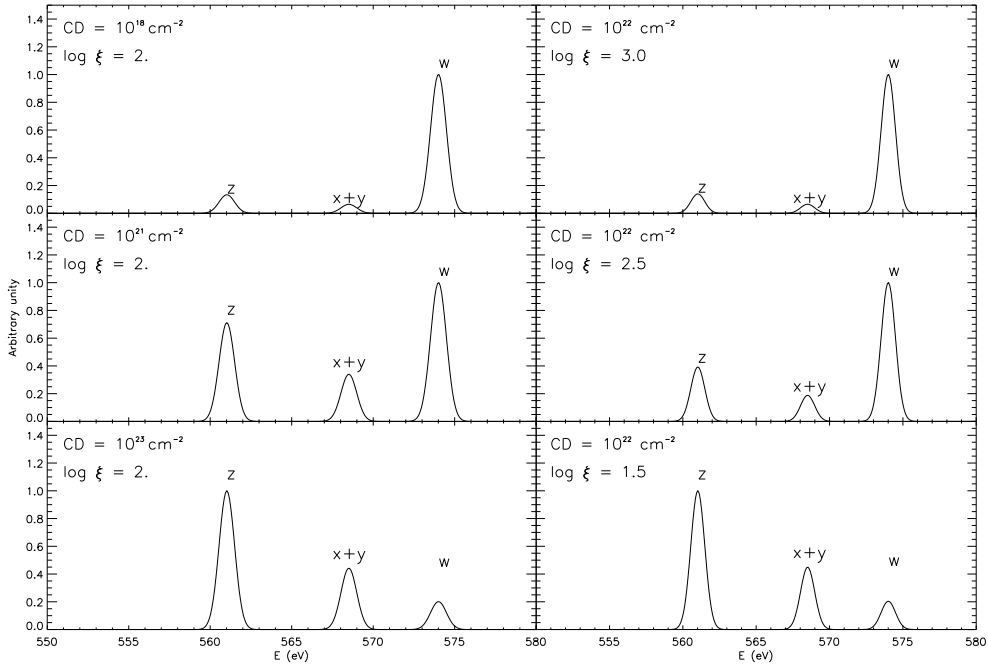,width=13.cm}
\caption{Spectrum showing the OVII triplet for several 
values of the column density and of the ionization parameter. 
The influence of an increasing column density at a given ionization 
parameter is shown on the left panels, and the influence of an 
increasing ionization parameter at a given column density  is shown 
on the right panel.}
  \label{Spectre_raieO7_n7}
\end{center}
\end{figure*}

\begin{figure*}
\begin{center}
\hspace{0.5cm}\psfig{figure=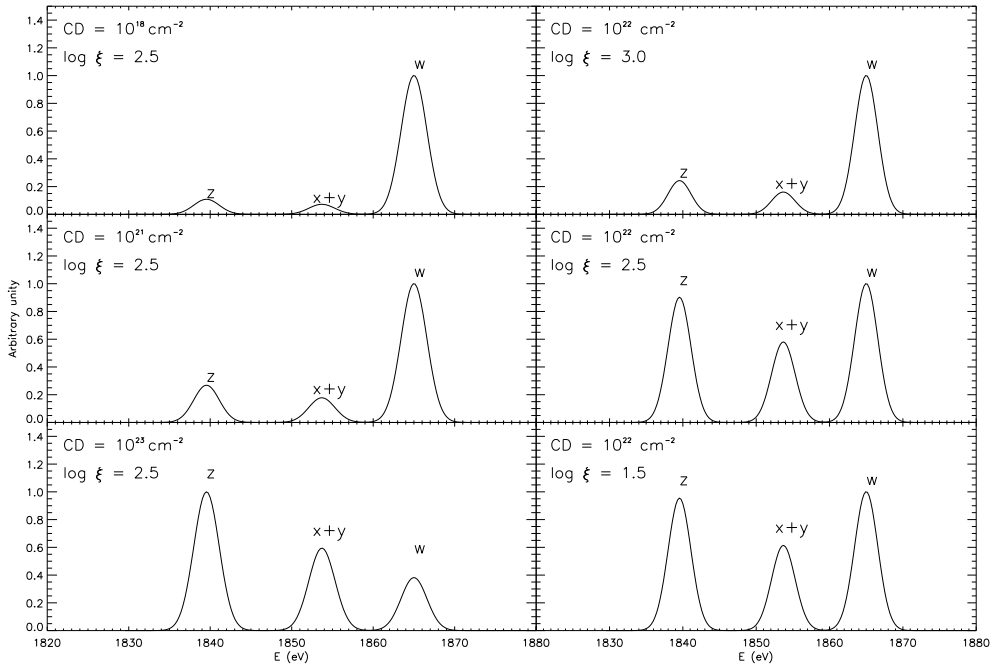,width=13.cm}
\caption{Same caption as Fig. \ref {Spectre_raieO7_n7} but for the 
SiXIII 
triplet. }
  \label{Spectre_raieSi13_n7}
\end{center}
\end{figure*}
\vspace{0.5cm}

   \begin{figure}
\begin{center}
\psfig{figure=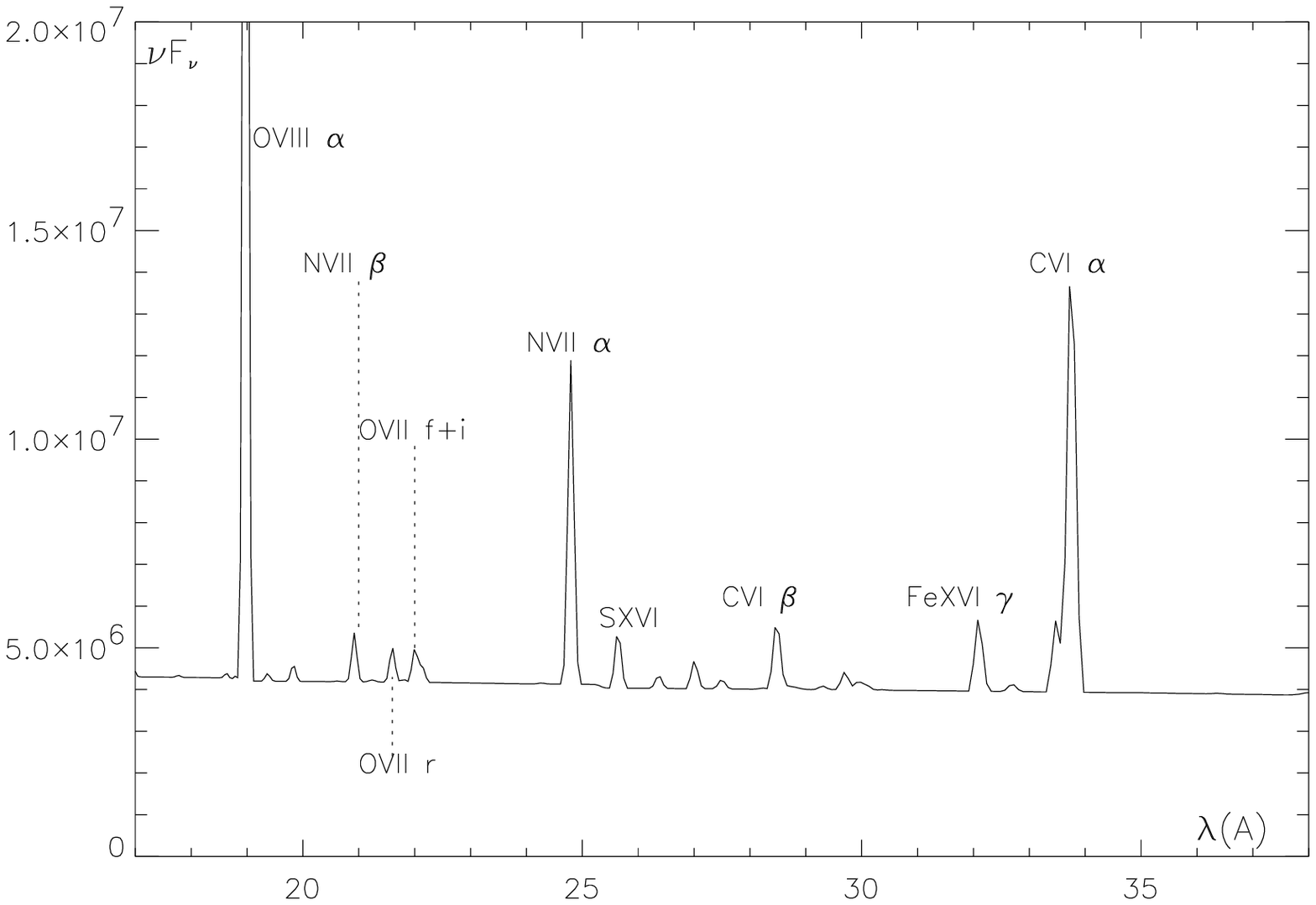,width=9.5cm,height=4.7cm}
\psfig{figure=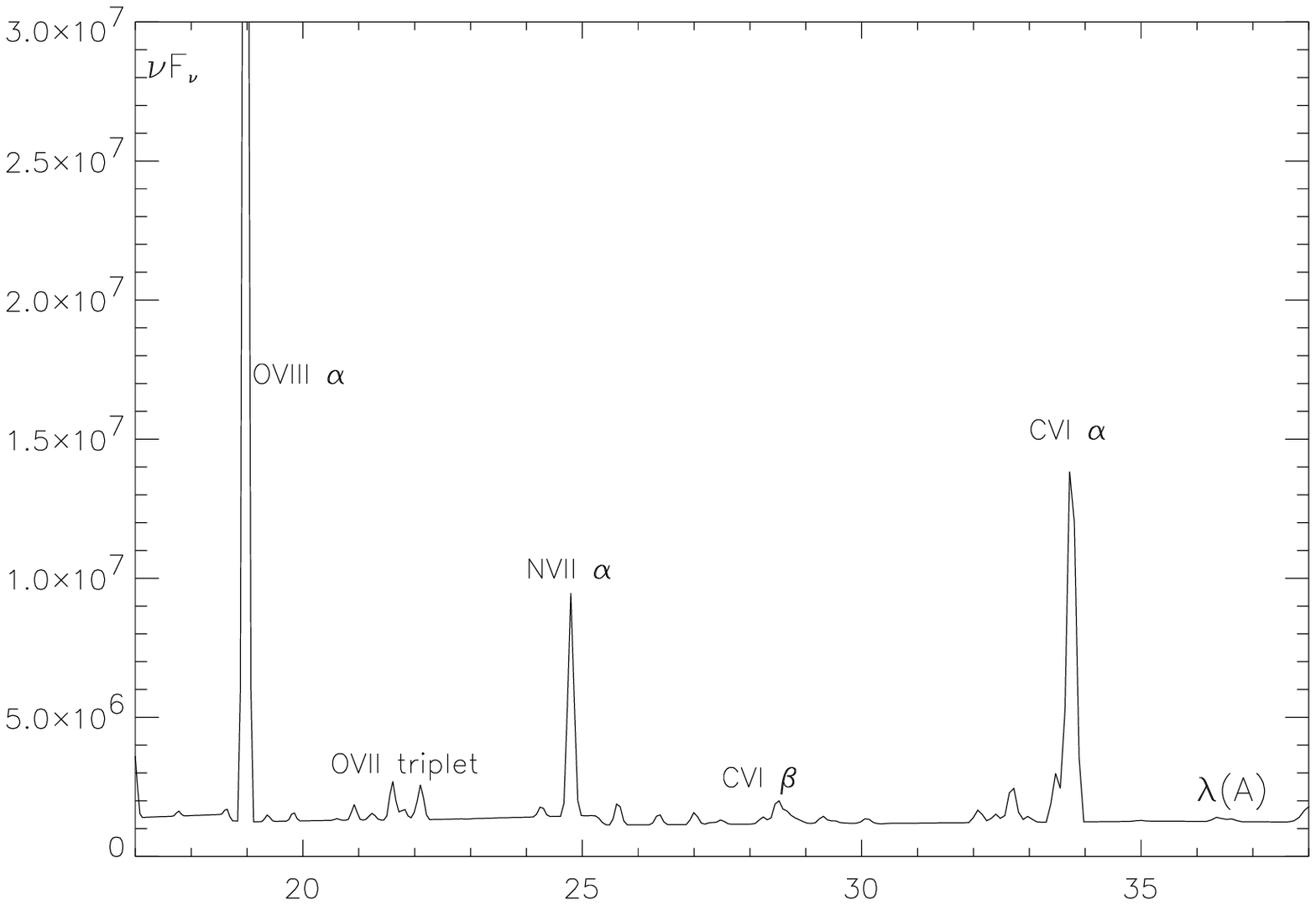,width=9.5cm,height=4.7cm}
\psfig{figure=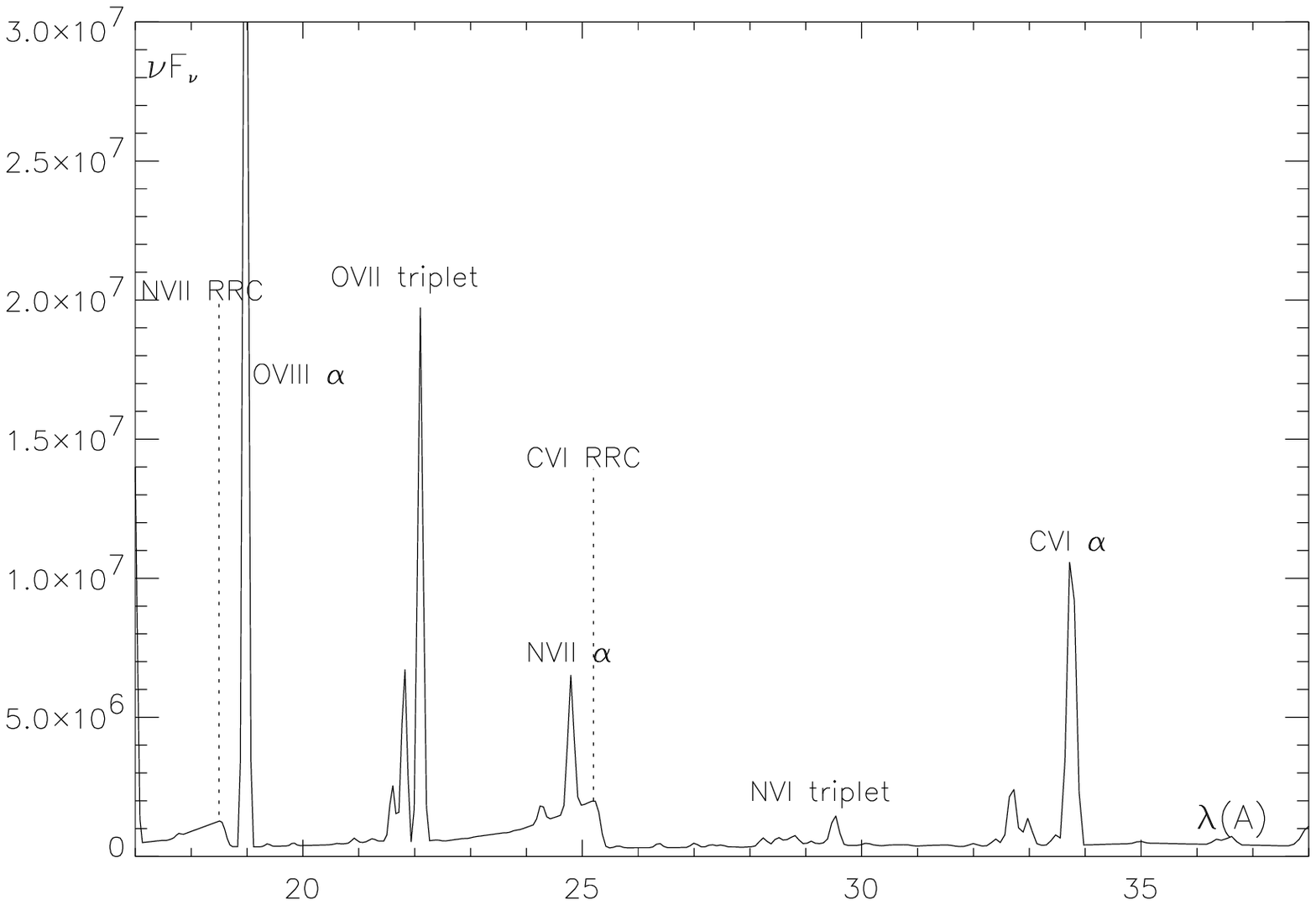,width=9.5cm,height=4.7cm}
\psfig{figure=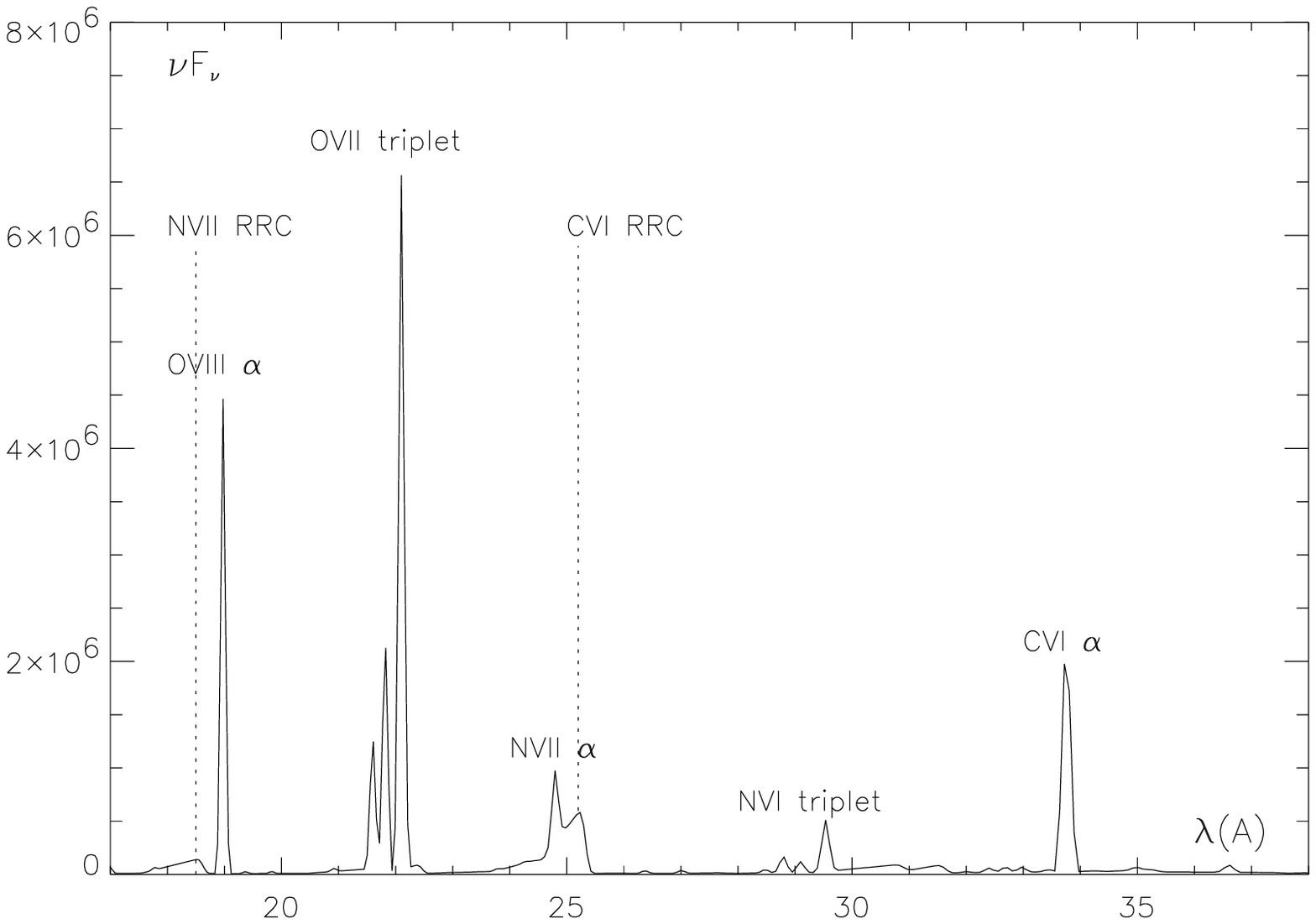,width=9.5cm,height=4.7cm}
\psfig{figure=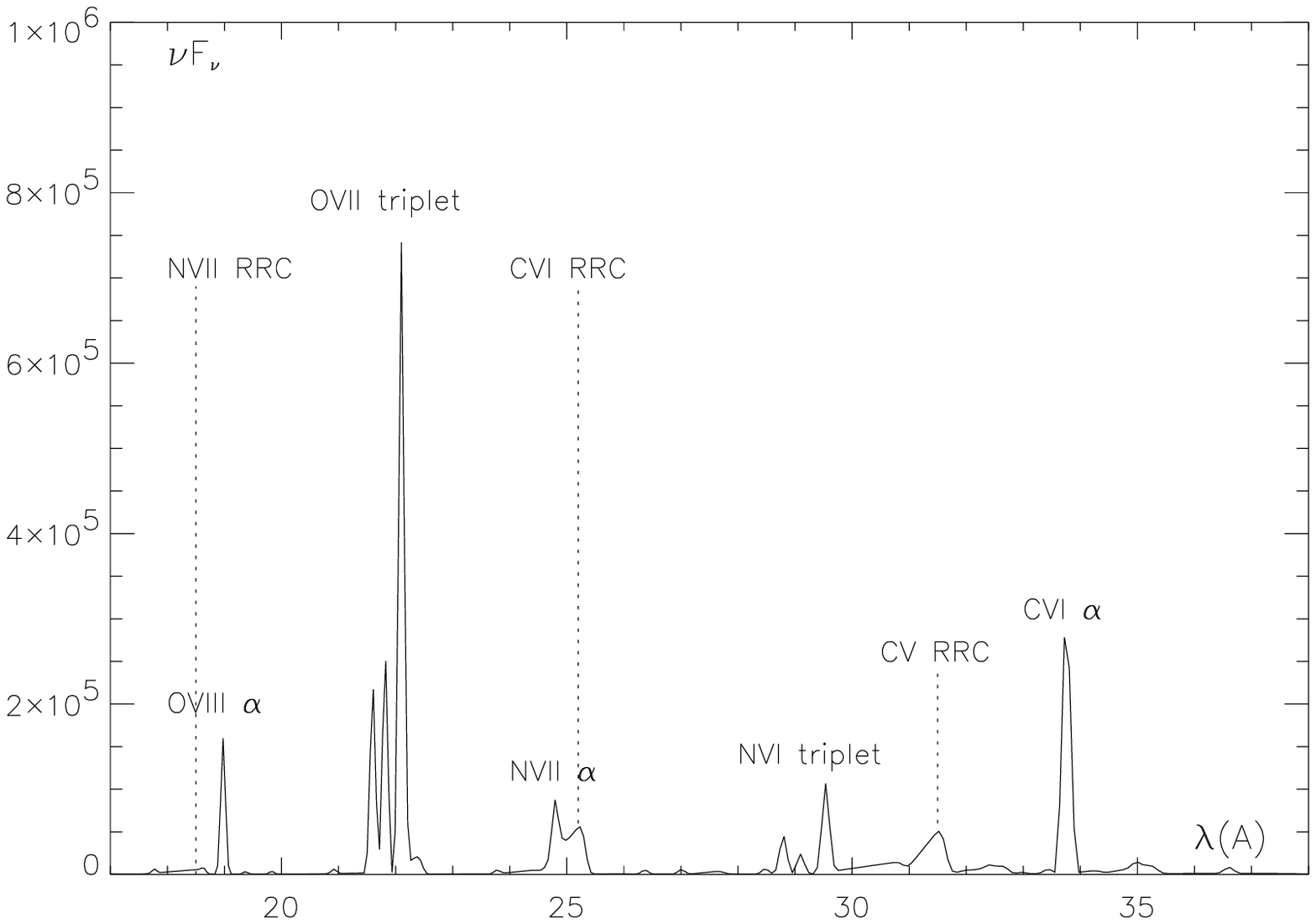,width=9.5cm,height=4.7cm}
\caption{A few examples of reflected spectra: from top to bottom: 
1. AGN continuum,
$n_{H}=10^{7}$ 
cm$^{-3}$, $CD=10^{23}$ cm$^{-2}$, and $\xi_{eq}=2000$; 2:  AGN 
continuum,
$n_{H}=10^{7}$ 
cm$^{-3}$, $CD=10^{23}$ cm$^{-2}$, and $\xi_{eq}=600$; 3:  AGN 
continuum,
$n_{H}=10^{7}$ 
cm$^{-3}$, $CD=10^{23}$ cm$^{-2}$, and $\xi_{eq}=200$; 4:  standard 
continuum,
$n_{H}=10^{7}$ 
cm$^{-3}$, $CD=10^{22}$ cm$^{-2}$, and $\xi=30$; 5: standard 
continuum, $n_{H}=10^{7}$ 
cm$^{-3}$, $CD=10^{21}$ cm$^{-2}$, and $\xi=10$. }
  \label{spectres}
\end{center}
\end{figure}

\begin{acknowledgements}
 We deeply aknowledge Franck Delahaye for providing us some 
collisional data for the OVII ion and Marie-Christine Artru for many 
enlightning discussions on atomic data.
\end{acknowledgements}

\end{document}